\documentclass[10pt,journal,compsoc,screen]{IEEEtran}

\ifCLASSOPTIONcompsoc
  \usepackage[nocompress]{cite}
\else
  \usepackage{cite}
\fi

\usepackage{mathtools}
\usepackage{booktabs} 
\usepackage{colortbl}
\usepackage{color}
\usepackage{xspace}
\usepackage{url}
\usepackage{balance} 
\usepackage{tabulary}
\usepackage{array}
\usepackage[ruled,lined,linesnumbered,vlined,algo2e]{algorithm2e}
\usepackage{subfig}
\usepackage{listings}
\usepackage{lipsum}
\usepackage{hyperref}
\usepackage{graphicx}
\graphicspath{{fig/}}
\usepackage[table,xcdraw]{xcolor}
\usepackage{amsmath,amsthm,amsfonts,amssymb,bm}
\usepackage[T1]{fontenc}
\usepackage[ansinew]{inputenc}
\usepackage{extarrows}
\usepackage{booktabs}
\usepackage{multirow}
\usepackage{colortbl}
\usepackage{tabularx}
\usepackage{enumitem}
\usepackage{verbatim}
\usepackage[numbers,sort&compress]{natbib}
\usepackage[justification=centering]{caption}
\usepackage{pifont}
\newcommand{\cmark}{\ding{51}}%
\newcommand{\xmark}{\ding{55}}%
\usepackage{makecell,multirow,diagbox,rotating}
\usepackage{longtable}
\usepackage{etoolbox} 
\usepackage{listings}
\usepackage{algorithm}
\usepackage{algpseudocode}
\usepackage{amsmath}
\usepackage[table,xcdraw]{xcolor}
\usepackage{tcolorbox}

\usepackage{enumitem}
\usepackage{wasysym}
\usepackage{tikz}
\usetikzlibrary{calendar,fpu}

\tikzset{
    moon colour/.style={
        moon fill/.style={
            fill=#1
        }
    },
    sky colour/.style={
        sky draw/.style={
            draw=#1
        },
        sky fill/.style={
            fill=#1
        }
    },
    southern hemisphere/.style={
        rotate=180
    }
}

\makeatletter
\pgfcalendardatetojulian{2010-01-15}{\c@pgf@counta} 
\def\synodicmonth{29.530588853}
\newcommand{\moon}[2][]{%
    \edef\checkfordate{\noexpand\in@{-}{#2}}%
    \checkfordate%
    \ifin@%
        \pgfcalendardatetojulian{#2}{\c@pgf@countb}%
        \pgfkeys{/pgf/fpu=true,/pgf/fpu/output format=fixed}%
        \pgfmathsetmacro\dayssincenewmoon{\the\c@pgf@countb-\the\c@pgf@counta-(7/24+11/(24*60))}%
        \pgfmathsetmacro\lunarage{mod(\dayssincenewmoon,\synodicmonth)}
        \pgfkeys{/pgf/fpu=false}
    \else%
        \def\lunarage{#2}%
    \fi%
    \pgfmathsetmacro\leftside{ifthenelse(\lunarage<=\synodicmonth/2,cos(360*(\lunarage/\synodicmonth)),1)}%
    \pgfmathsetmacro\rightside{ifthenelse(\lunarage<=\synodicmonth/2,-1,-cos(360*(\lunarage/\synodicmonth))}%
    \tikz [moon colour=white,sky colour=black,#1]{
        \draw [moon fill, sky draw] (0,0) circle [radius=1ex];
        \draw [sky draw, sky fill] (0,1ex)
            arc (90:-90:\rightside ex and 1ex)
            arc (-90:90:\leftside ex and 1ex)
            -- cycle;
    }%
}

\usepackage{balance}
\usepackage{makecell}

\newcommand{\revise}[1]{\textcolor{red}{#1}}
\newcommand{\xian}[1]{\textcolor{violet}{#1}}

\AtBeginDocument{%
  \providecommand\BibTeX{{%
    \normalfont B\kern-0.5em{\scshape i\kern-0.25em b}\kern-0.8em\TeX}}}

\begin{document}

\title{A Systematic Assessment on Android Third-party Library Detection Tools}

\author{ Xian Zhan,
         Tianming~ Liu,
         Yepang Liu,
         Yang Liu,
         Li Li,
         Haoyu~Wang,
         Xiapu~Luo
\IEEEcompsocitemizethanks{
\IEEEcompsocthanksitem Xian Zhan and Xiapu Luo are with Department of Computing, The Hong Kong Polytechnic University, Hong Kong, China.
\IEEEcompsocthanksitem Yepang Liu is with Department of Computer Science and Engineering, Southern University of Science and Technology, Shenzhen, China.
\IEEEcompsocthanksitem  Yang Liu is with School of Computer Science and Engineering, Nanyang Technological University, Singapore. 
\IEEEcompsocthanksitem {Haoyu Wang is with  School of Computer Science, Beijing University of Posts and Telecommunications, China}
\IEEEcompsocthanksitem Tianming Liu and Li Li are with Monash University, Australia.
}
\thanks{Xiapu Luo (csxluo@comp.polyu.edu.hk) is the corresponding author;}
}

\markboth{Journal of \LaTeX\ Class Files,~Vol.~xx, No.~xx, xx~2020}%
{Shell \MakeLowercase{\textit{et al.}}: titleXXX}


\IEEEtitleabstractindextext{%
\begin{abstract}
Third-party libraries (TPLs) have become a significant part of the Android ecosystem. Developers can employ various TPLs to facilitate their app development. Unfortunately, the popularity of TPLs also brings new security issues. For example, TPLs may carry malicious or vulnerable code, which can infect popular apps to pose threats to mobile users. Furthermore, TPL detection is essential for downstream tasks, such as vulnerabilities and malware detection. Thus, various tools have been developed to identify TPLs. However, no existing work has studied these TPL detection tools in detail, and different tools focus on different applications and techniques with performance differences.
A comprehensive understanding of these tools will help us make better use of them.


To this end, we conduct a  comprehensive empirical study to fill the gap by evaluating and comparing all publicly available TPL detection tools based on six criteria: accuracy of TPL construction, effectiveness, efficiency, accuracy of version identification, resiliency to code obfuscation, and ease of use. 
Besides, we enhance these open-source tools by fixing their limitations, to improve their detection ability. Finally, we build an extensible framework that integrates all existing available TPL detection tools, providing an online service for the research community. We release the evaluation dataset and enhanced tools. According to our study, we also present the essential findings and discuss promising implications to the community; e.g., 1) Most existing TPL detection techniques more or less depend on package structure to construct in-app TPL candidates. However,  using package structure as the module decoupling feature is error-prone. We hence suggest future researchers using the class dependency to substitute package structure. 2) Extracted features include richer semantic information (e.g., class dependencies) can achieve better resiliency to code obfuscation.
3) Existing tools usually have a low recall; that is because previous tools ignore some features of Android apps and TPLs, such as the compilation mechanism, the new format of TPLs, TPL dependency. Most existing tools cannot effectively find partial import TPLs, obfuscated TPLs, which directly limit their capability. 4) Existing tools are complementary to each other; we can build a better tool via combining the advantages of each tool.
We believe our work provides a clear picture of existing TPL detection techniques and also gives a road-map for future research.

\end{abstract}

}



\maketitle

\section{Introduction}
\label{sec:Introduction}
\subsection{Motivation}
\IEEEPARstart{N}owadays, Android applications (apps) occupy an irreplaceable dominance in the app markets~\cite{statista} and will continuously hit the new height~\cite{appfuture}. Along with the thriving of Android apps is the emerging of countless third-party libraries (TPLs). When app developers implement their own apps, they usually realize some {common} functionalities by integrating various TPLs (to avoid reinventing the wheel), such as advertisements, social networking, analytics, etc. 
Prior research~\cite{PEDAL2015MobiSys} has shown that about 57\% of apps contain third-party ad libraries. Wang et al.~\cite{Wang2017ICSE-C} also revealed that more than 60\% of the code in an Android app belongs to TPLs. 
However, this situation could bring new security threats. When the malicious TPLs {and vulnerable TPLs} are integrated into popular apps, they may quickly infect a large number of mobile devices.
Besides, TPLs as noises could affect the detection results of repackaging detection ~\cite{xian2019saner}, malware detection~\cite{Lili2016SANER}, etc. Thus, research on TPL detection targeting the Android platform continues to emerge.

{Generally, there are two ways to identify TPLs. The first one is the whitelist-based approach, which uses the package name to find the in-app TPLs.
The second one is the signature-based approach, which extracts features from TPLs to identify them. In the beginning, most repackaged app detection~\cite{Juxtapp13DIMVA,DroidMOSS12CODASPY,ViewDroid14,ResDroid2014shao} and malware detection {techniques}~\cite{MassVet2015chen} adopt whitelist to filter out TPLs because whitelist-based approach is simple and easy to implement. However, 
the whitelist-based method is not resilient to package renaming. A recent study showed that more than 50\% of the inspected Android TPLs are protected by obfuscation techniques~\cite{Lin2014soups}, which dramatically decreases the effectiveness of the whitelist-based method. Besides, the whitelists cannot cover all TPLs, especially newly-emerged ones.}

In order to improve the detection performance, various research ~\cite{AdDetect2014ISSNIP,PiggyApp13CODASPY,LibRadar2016ICSE,LibD2017ICSE,libscout2016ccs,LibSift2016soh,ORLIS2018MOBILESoft} have emerged one after another and adopt different techniques to identify in-app TPLs. 
However, there are still many mysteries about {previous TPL detection tools}. For instance,
the advantages and disadvantages, usage scenarios, performance, and resiliency to code obfuscation of these tools are still unclear. 
Besides, no unified dataset is available to quantitatively evaluate them without bias. {Undoubtedly, it is necessary to conduct a thorough comparison of previous TPL detection tools. Such a study can help users provide the most suitable tool for a specific application scenario; help researchers understand limitations in current TPL detection techniques; propose useful insights for future research.} Both researchers and developers can know the current status of TPL detection, and future researchers may be enlightened by reading such a research paper.
%
To this end, we attempt to conduct a comprehensive and fair comparison of these state-of-the-art TPL detection tools on an unbiased dataset.
We evaluate them using six metrics: in-app TPL candidate construction accuracy, effectiveness, efficiency/scalability, exact version identification capability, code obfuscation-resilience capability, and ease of use. Via such an evaluation, we can thoroughly understand the performance of existing tools from different perspectives.
\vspace{-2ex}
\subsection{Contributions}
By investigating the six aspects of these tools, we attempt to achieve five goals in this study: (1) understand the capabilities and usage scenarios of different TPL detection tools. (2) conclude a better-optimized scheme to guide future work or help developers implement better tools. (3) release a well-designed benchmark that enables to evaluate current and future tools. {This dataset includes the possible situations of imported TPLs and covers the challenges during the TPL detection. To evaluate the accuracy of different tools in module decoupling, we collect a set of apps including the TPLs, and these datasets cover various situations in TPL construction. To evaluate the effectiveness and efficiency, we collect 221 Android apps from Google Play and their corresponding 59 unique TPLs with 2,115 version files.
To verify the version identification accuracy, we collect two different datasets that include the two sets of open-source apps from F-Droid with their corresponding in-app TPLs and version files. To evaluate the resilience to code obfuscation,
We also build a dataset that includes obfuscated apps with various obfuscation techniques by different obfuscators.} (4) enhance existing tools to achieve better performance based on our study;
(5) integrate these publicly available tools as an online service, which provides the detection results of different TPL detection tools to users. The purpose of integrating multiple tools as an online service is to make up for the shortcomings of each tool.
In summary, our main contributions are as follows:

\begin{itemize}

\item We are the first to conduct a systematic and thorough comparison of existing TPL detection tools by using six different metrics. {Based on our analysis, we find the advantages and disadvantages of current research and provide essential findings and present the potential challenges in this direction. We give suggestions on tool selection under different application scenarios; propose a better idea on tool implementation and provide useful insights for future research in TPL detection.} 

\item {We are the first one to construct a comprehensive benchmark to evaluate TPL detection tools from different aspects. Based on these datasets, we can systematically evaluate the performance of TPL detection tools from different aspects, to let readers thoroughly understand this direction.
}


\item We attempt to improve the practical values of these publicly available tools. We fixed bugs and {address the limitations} of some publicly available tools for better performance. We also implement an extensible framework that integrates all existing TPL detection tools to provide an online service to users. All the related code and dataset and detailed evaluation results can be found on our website\footnote{https://sites.google.com/view/libdetect/}, which can also facilitate the future research.





\end{itemize}

This paper makes a substantial extension of a previous conference paper~\cite{libdetect2020ASE}, which was published at the 35th IEEE/ACM International Conference on Automated Software Engineering (ASE 2020). The major changes in this version are summarized as follows:

\begin{itemize}
\item A new section (cf. Sec.~\ref{Sec:challenge}) is added to elaborate primary challenges that are usually ignored by previous research in TPL detection. We conduct a thorough analysis on the various in-app TPLs, and conclude TPLs imported methods, dependencies of TPLs, obfuscated types on TPLs, various and complicated package structures of in-app TPLs, etc.

\item We give more detailed introduction to TPL detection tools. Meanwhile, we add three recent tools, i.e., LibExtractor~\cite{LibExtractor2020wisec}, LibRoad~\cite{LibRoad2020TMC} and OSSPoLICE~\cite{OSSPOLICE2017CCS}.
We compare the new tools as well and present a more detailed and complete summary for each tool based on the detection process (cf. Sec.~\ref{sec:comparison} \& Sec.~\ref{sec:evaluation}). 


\item We add another two new metrics (i.e., C1 and C4) to evaluate state-of-the-art TPL identification tools in terms of the accuracy of in-app TPL candidate construction and version identification capability.
The evaluation design can refer to Sec.~\ref{sec: empirical study} and the experimental setups and results can refer to Sec.~\ref{sec:evaluation:C3}.


\item To conduct the two new experiments, we also build corresponding new well-designed datasets. For C1, we try to verify their accuracy of TPL candidates construction. {Thus, we designed a dataset that contains almost the challenging cases that may be encountered in TPL candidates construction in the real-world.
}
For C4, we collect two datasets that include the apps and their used TPLs with corresponding version files to verify the accuracy of version identification. {For the first dataset, we collected TPLs with large code differences among different versions. For the second dataset, we collected TPLs with small differences among different versions.} 


\item We also give more detailed analysis on effectiveness and efficiency on these publicly available tools. Besides, we also add a new metric (the time complexity) and new experiment to evaluate the efficiency of these state-of-the-art tools.


\item We also add the implications and provide ideas about how to design a more effective TPL detection tool for future researchers (cf. Sec.~\ref{sec:discussion}). 

\end{itemize}

\section{Preliminary}
\label{sec:Background}

\subsection{Android Third-party Library}
\label{sec:background:tpl}

Third-party library (TPL) provides developers with various standalone functional modules, which can be integrated into host apps with the help of many development tools (e.g., Android Studio, Eclipse) to speed up the development process. Since current Android TPL detection tools that we compared in this paper only consider Java libraries, we do not discuss the native libraries (C/C++) here. The Java libraries are usually published as ``.jar'' or ``.aar'' files. The ``.aar'' format file is a kind of exclusive TPLs for Android apps, which usually provides UI-related libraries or game engine libraries. The ``.jar'' files consist of class bytecode files, while ``.aar'' files include both class bytecode files and other Android-specific files such as the manifest file and resource files. 
Most TPL files can be found/downloaded/imported from maven repository~\cite{maven}, Github~\cite{github}, and Bitbucket~\cite{bitbucket}. 
A TPL can be uniquely defined by its ``group ID'', ``artifact ID'', and version number~\cite{maven}.
In Android app development, if an app uses TPLs, the app code can be divided into two parts, the logic module from the host app (i.e., primary) and the supplementary modules (i.e., non-primary) from TPLs \cite{LibSift2016soh,AdDetect2014ISSNIP}. TPL detection aims to identify TPLs in non-primary modules. {Developers also have different ways to import TPLs, such as importing from the source code, importing from the TPL repositories with the help of development IDE.}


\subsection{TPL Detection Process}

\begin{figure}[t]
    \centering
    \includegraphics[scale=0.42]{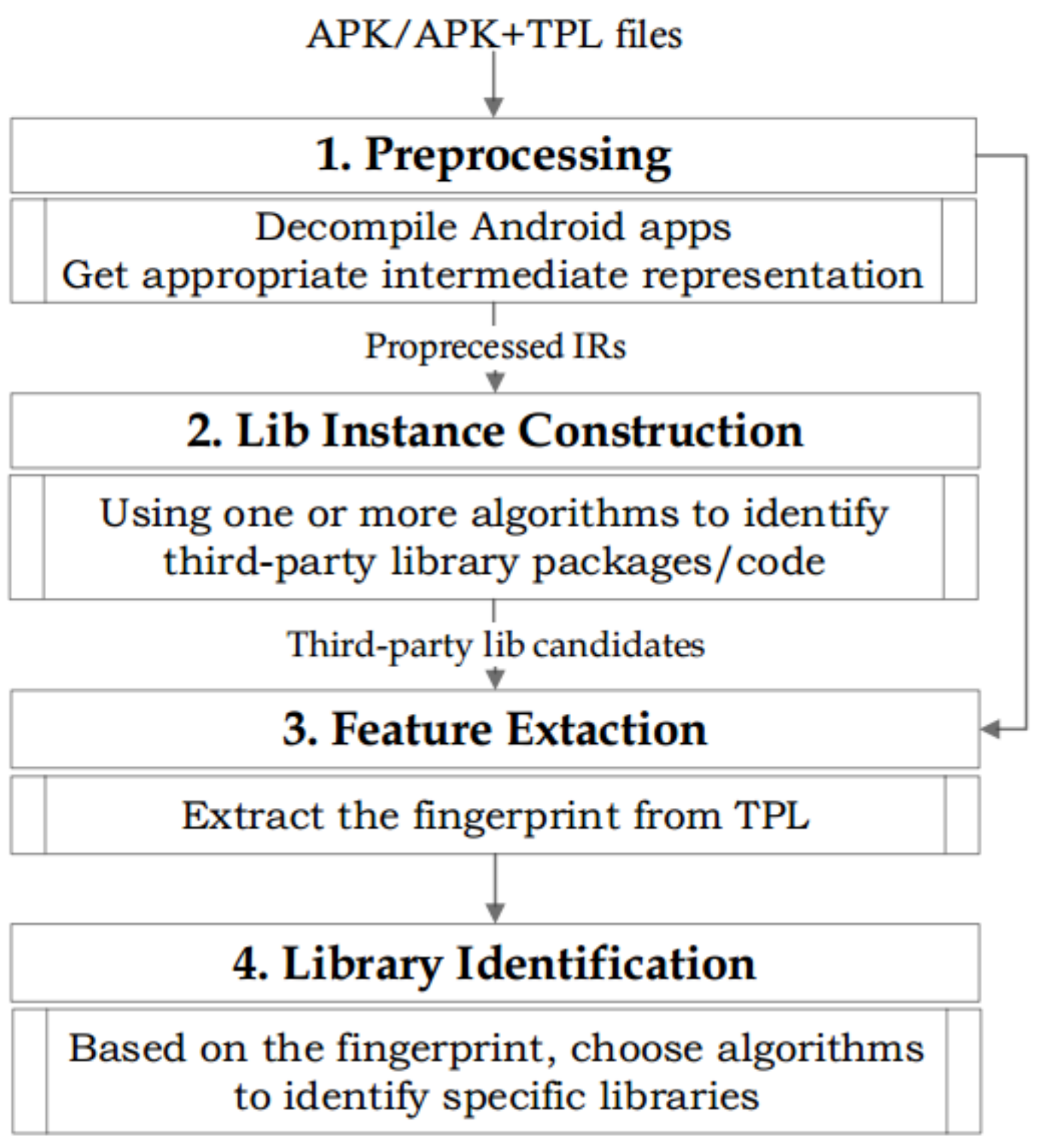}\\
    \vspace{-1.6mm}
    \caption{Typical process of TPL detection}
    \label{fig:process}
    \vspace{-3ex}
\end{figure}

According to our analysis, we find that existing state-of-the-art detection techniques usually unfold in four steps (see Fig.~\ref{fig:process}), which are elaborated below.

\noindent \textbf{\textit{Step1: }Preprocessing.}
Researchers usually first decompile apps by applying reverse-engineering tools and then get the appropriate intermediate representations (IRs) in this stage. 
Based on different IRs, the corresponding techniques can be applied to facilitate the following steps.




\noindent \textbf{\textit{Step 2: }Library Instance Construction.}
{As can be seen from Fig.~\ref{fig:structure}, if an Android app includes TPLs, we can divide the app into two modules: the primary module that is composed of code from host app, the non-primary module that may consist of code from one or more TPLs.
This step aims to separate TPLs from the host app and then
find the boundaries of each TPL candidate from the non-primary module. This step is also called module decoupling.}

\noindent \textbf{\textit{Step 3: }Feature Extraction.}
In this step, researchers choose an appropriate transformation method to generate TPL features. The extracted code features must uniquely represent different TPLs.
Existing tools usually extract features such as Android APIs, control flow graphs, and variant method signatures to represent TPLs.


\noindent \textbf{\textit{Step 4:} Library Instance Identification.}
Existing identification methods can be classified into two types based on the differences in TPL feature database construction: \textit{clustering-based} method and \textit{similarity comparison} method. {Note that, existing classification-based systems are all binary classification; they only can distinguish the ad and non-ad libraries rather than identify different TPLs.}
The clustering-based method usually depends on sophisticated module decoupling techniques. 
This method first needs to filter out the primary module (i.e., code of the host app) and then cluster non-primary modules with similar features together. The modules in one cluster are considered as a TPL. 
The similarity comparison method requires collecting TPL files in advance and then uses the same feature extraction algorithm to generate the TPL signatures to build a pre-defined TPL signature database.
By comparing the similarity of the features between the collected TPLs and the in-app TPL candidates, in-app TPLs can be identified.

\subsection{Challenges in TPL Detection}
\label{Sec:challenge}
\begin{figure}[t]
    \centering
    \includegraphics[scale=0.48]{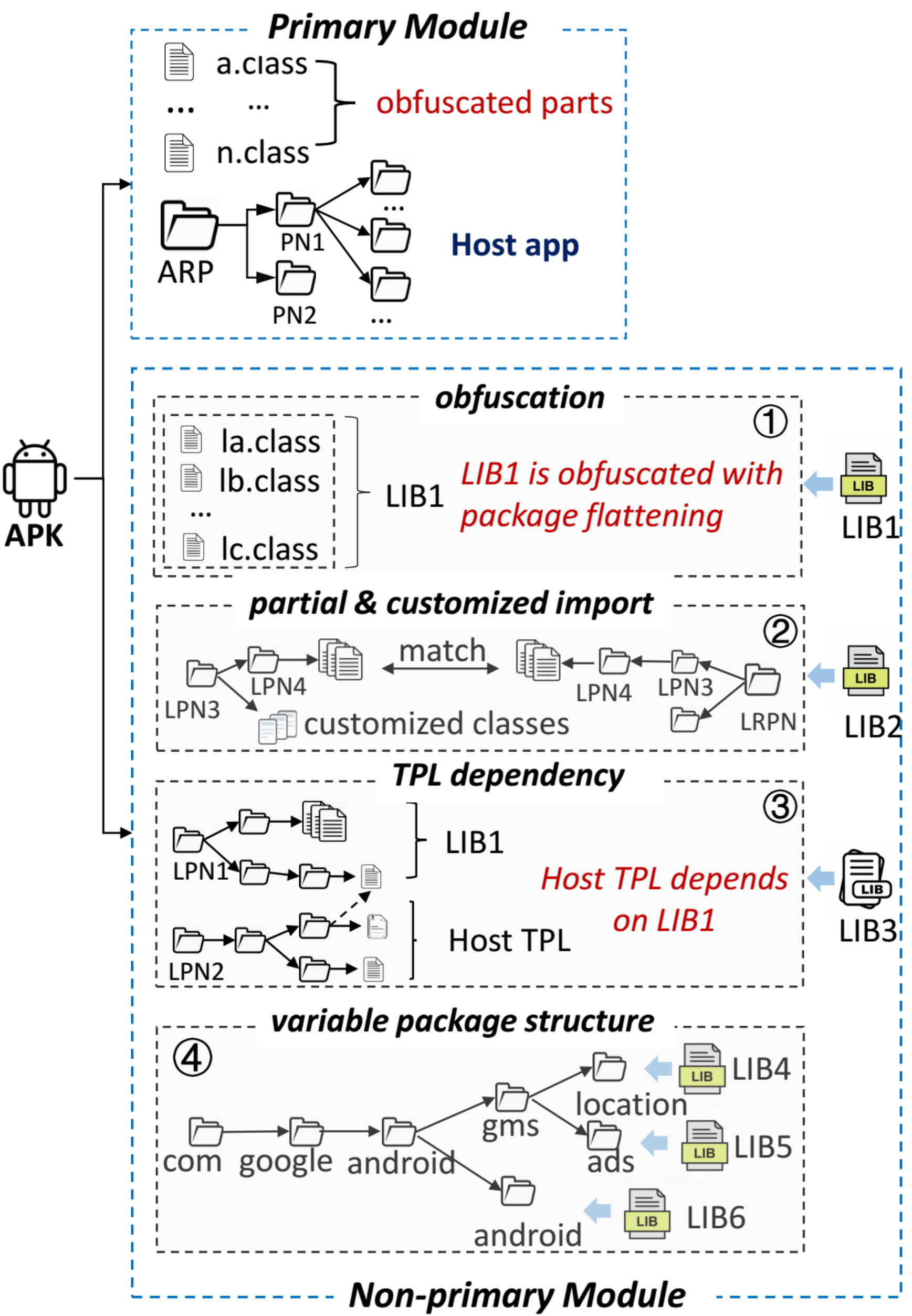}\\
    \vspace{-1.mm}
    \caption{The composition of a decompiled APK}
    	\begin{center}
		\textit{ARP: app root package; PN: package name, LPN: library package name}
	\end{center}
    \label{fig:structure}
    \vspace{-4ex}
\end{figure}

After we have introduced the TPL detection process, we now discuss the challenges in TPL detection.
Based on our analysis, the challenges of TPL detection can be summarized as four points: 1) code obfuscation, 2) various imported modes, 3) TPL dependency, and 4) complicated package structures. We give a motivating example of the composition of a decompiled APK in Fig.~\ref{fig:structure} that illustrates the aforementioned challenges; we elaborate on these challenges below.

\vspace{-1ex}

\subsubsection{Code Obfuscation Strategies}
\label{sec:challenge:obfuscation}
Code obfuscation~\cite{obfuscation} is often used to protect software against reverse engineering. There are many obfuscators (i.e., obfuscation tools) such as Allatori~\cite{Allatori}, DashO~\cite{Dasho}, Proguard~\cite{Proguard} helping developers obfuscate their apps and TPLs.
Some obfuscation techniques can hide the actual logic of the apps as well as the used libraries. TPL detection tools generate code feature to represent each TPL; however, these code obfuscation techniques can lead to the modification of these code features, resulting in mis-identification.
The commonly-used Android app obfuscation strategies are introduced as follows.


\noindent \textbf{Identifier Renaming}, which renames identifiers into meaningless characters such as ``a'' and ``b'' or hash values. The identifiers include the package names, class names, the method names and the field names, etc~\cite{renaming}.

\noindent \textbf{String Encryption}, which usually adopts encryption algorithms to protect sensitive information such as telephone or email. After encryption, the sensitive strings defined in the source code are encrypted to meaningless strings~\cite{string_encryption}.

\noindent \textbf{Package Flattening}, which modifies the package hierarchy structure by moving the files from one folder to another. Different obfuscators can flatten the structure to varying degrees. Note that, package structure can change the inter and intra hierarchy structure.
Sometimes the whole package structure can be removed, and all the files are put into the root directory of apps~\cite{flattening,ORLIS2018MOBILESoft}.




\noindent \textbf{Dead Code Removal}, which deletes unused code and preserves the functionalities invoked by the host app~\cite{coderemoval}.

\noindent \textbf{API hiding}, which exploits Java reflection and dynamic class loading techniques to hide APIs. Many adversaries usually adopt this technique to hide malicious behaviors and it is difficult for static analysis to find them. The common usage is to add the annotation @hide at the front of nonpublic APIs in TPLs~\cite{Dong2018obfuscation}.

\noindent \textbf{Control Flow Randomization}, which modifies the Control Flow Graph (CFG) without changing the actual execution tasks, e.g., inserting redundant control flow or flattening control flows~\cite{CFR}. 

	

\noindent \textbf{Dex Encryption}, 
which allows developers to encrypt the whole DEX file. It can encrypt user-defined functions as well as Android components such as Activities and Services. The protected classes would be removed from the original classes.dex files, thus, cannot be obtained by reverse-engineering tools~\cite{dex_encryption}. 
	
\noindent \textbf{Visualization-based Protection}, 
which translates the code into a stream of pseudo-code bytes that is hard to be recognized by the machine and human. 
To defend against it, the app should be in a specific execution during runtime, where the pseudo-code is interpreted one by one, and gradually restored to the original code, then it can be executable~\cite{Dong2018obfuscation}.

{As shown in Fig.~\ref{fig:structure}, a part of classes from the host app and LIB1 are obfuscated. The class files were renamed by using a meaningless character such as `a', `b'. Besides, things could be worse. If this app is obfuscated with the package flattening technique, the renamed classes in the root directory may be from the package tree of the host app or TPLs. Package flattening obfuscation can change the inter and inter package structure and even can remove the whole package tree. Therefore, the challenge is how to distinguish the obfuscated classes from the host app or LIB1.}

\subsubsection{Various Import Modes}
\label{sec:import mode}

{When developers use TPLs in app development, the imported mode usually can be divided into three types: 1) complete import, which means all the packages and classes of a TPL are entirely integrated into one app. 2) partial import, only import a part of functions (partial classes and packages) of a TPL. Developers can use obfuscation tools (e.g., Proguard) to remove some no invoked code of a TPL in an app, which can decrease the size of published apps. Without a doubt, that will affect the detection performance since the original library is different from the imported one. 3) customized import. If developers find the invoked TPL cannot implement some functions, developers can modify the code of TPL to meet their requirements. Some code logic could be change during this processing, causing the methods and classes to be different from the original TPLs.
As shown in Fig.~\ref{fig:structure}, the partial import TPL can only match a part of the original TPL. 
To some extent, partial and customized import modes may affect the detection performance. Thus, how to solve this problem is a TPL detection tool that needs to consider.}

\subsubsection{TPL Dependency}
\label{sec:challenge:TPL_dependency}
Apart from apps, many TPLs are also built on other TPLs. We investigated about 50K TPLs and found about 47\% of them are dependent on other TPLs. For example, the popular TPL ``Google Guava''~\cite{guava} ``Jackson Databind''~\cite{jackson-databind} are widely used by more than 24K and 16k TPLs, respectively.

{As can be seen the case 3 in the Fig~\ref{fig:structure}, for TPLs with dependency, the package structure may be independent but their code has logic links, the multiple independent package trees and files consist of a complete TPL. 
If a TPL detection tool just adopts the package tree to construct in-app TPL instances, a complete TPL may be separated into several parts.} 
{Meanwhile, we should pay attention to that Java has different package techniques~\cite{uber}, it can make the Java applications with different package types (i.e., ``skinny'', ``thin'', ``Fat/Uber'', and ``Hollow'').
Different package types can let the published version include or not include the dependencies. For example, if a TPL adopts the ``skinny'' package technique, the published TPL file will not include the dependencies. Only when an app import this TPL during the compilation process, the dependencies of this TPL will be loaded into the app, which leads to the published app including the host TPL and its third-party dependencies. That means the code of the published original TPL is different from the code of in-app TPL. Undoubtedly, that brings extra challenges to TPL detection. To accurately identify TPLs, developers should not only identify TPL dependencies, but also pay attention to package technologies.}

\subsubsection{Variable Package Structures}
\label{sec:challenge:package}

When we consider a single original TPL file, it may have a well-organized and complete package hierarchy structure. However, if we import a TPL into an Android app, things will be more complicated. If we use the package name and structure to identify a TPL, it may create some mistakes. In some cases, a TPL may not have any package structures (code obfuscation (package flattening technique)). In some cases, a complete TPL may have multiple parallel package structures (TPL dependency). In some cases, multiple independent TPLs may share the same root package. In some cases, the same TPL with different versions may have different package structures and names.

As shown the case 4 in Fig~\ref{fig:structure}, the ``Play Services Location'', ``Play services Ads'' and ``Google Android Library'' are three different TPLs. They are represented as LIB4, LIB5, and LIB6 in Fig.~\ref{fig:structure}, respectively. Nevertheless, we can see these TPLs share the same root package \textit{``com.google.android''}. The LIB4 and LIB5 share the same root package structure \textit{``com.google.android.gms''}. This situation is very prevalent. If these TPLs belong to the same group or are developed by the same team,
these different TPLs may have the same root package name.
Apart from the above-mentioned situations, some companies or organizations also published some TPLs with different granularity. For example, the Google Android Design support library, to fulfill different developers requirement, it published the coarse version~\cite{design} and some fine-granularity version files~\cite{design_color,design_resources}.
Besides, regarding the same TPLs with different versions, the package structure and name also can be mutant. For example, the TPL OkHttp, a popular HTTP client, before the version 3.0, the package name is ``com.squareup.okhttp'' while after the version 3.0, the package name become ``okhttp3''.
Without a doubt, the above-mentioned situations have brought great challenges to TPL detection. 
In the following sections, we will explore whether these state-of-the-art tools can handle aforementioned challenges well.

\vspace{-1ex}
\section{Related Work}
\label{sec:relatedwork}

Third-party library detection plays an important role in the Android ecosystem, such as malware/repackaging detection, where TPLs are considered as noises, thus should be filtered out. Most malicious/repackaged apps detection employed a whitelist-based method to detect TPL based on the package name.
Chen \textit{et al}.~\cite{chen2014achieving} collect 73 popular libraries as the whitelist to filter third-party libraries when detecting app clones.
Repackaging detection tools~\cite{DroidMOSS12CODASPY,Juxtapp13DIMVA,chen2014achieving} and malware detection tool~\cite{MassVet2015chen} also adopt the whitelist-based method to remove third-party libraries. 
However, such a method exists hysteresis and lacks robustness, which cannot cover all TPLs and finds emerging libraries, as well as obfuscated libraries.
%
To seek more effective approaches, various tools of TPL detection appear. This paper elaborates on these tools and evaluates them from different perspectives. Similar comparison studies are as follows.

\noindent \textbf{Android Testing Tool Comparison.}
Shauvik et al.~\cite{Testcompare2015ASE} compared the effectiveness of Android test input generation tools based on four aspects: ease of use, compatibility, code coverage, and fault detection ability. They reveal the strengths and weaknesses of different tools and techniques.
Zeng et al.~\cite{test2016FSE} also conducted an empirical study of various Android input generation tools and found Monkey could get the best performance. 
They also developed a new method to improve the code coverage of Monkey.
Kong et al.~\cite{kong2019Testing} reviewed 103 papers related to automated testing of Android apps. They summarized the research trends in this direction, highlighted the state-of-the-art methodologies employed, and presented current challenges in Android app testing. 
They pointed out that new testing approaches should pay attention to app updates, continuous increasing of app size, and the fragmentation problem in the Android ecosystem. Fan et al.~\cite{fan2018large,fan2018efficiently,su2020my} evaluated the effectiveness of both dynamic testing tools and static bug detection tools in Android apps, especially for Android-specific bugs.

\noindent \textbf{Clone App Detection Comparison.}
Li et al.~\cite{lili2019TSE} surveyed 59 state-of-the-art approaches of repackaged app detection, in which they compared different repackaging detection techniques and elaborated current challenges in this research direction. They found that current research on repackaging detection is slowing down. They also presented current open challenges in this direction and compared existing detection solutions. Besides, they also provided a dataset of repackaged apps, which can help researchers reboot this research or replicate current approaches.
Zhan et al.~\cite{xian2019saner} conducted a comparative study of Android repackaged app detection. They reproduced all repackaged app detection tools and designed a taxonomy for these detection techniques and then analyzed these techniques and compared their effectiveness. Finally, they listed the advantages and disadvantages of current techniques. 
Furthermore, Baykara et al.~\cite{repackaged_survey2018ISDFS} investigated malicious clone Android apps. They revealed potential threats that can affect users' experience and also provided potential solutions for these risks.

\noindent \textbf{Relevant Techniques.}
We can find that many clone detection techniques are used in TPL detection.
Early app clone detection techniques~\cite{DroidMOSS12CODASPY,ORLIS2018MOBILESoft,Juxtapp13DIMVA} usually use opcode as the feature and use hash algorithms to generate the fingerprints.
\textit{DroidMOSS}~\cite{DroidMOSS12CODASPY} extracts opcode as the fingerprints and employs a fuzzy hashing technique to localize the changing parts in the clone apps, whose idea is similar to that in \textit{ORLIS}~\cite{ORLIS2018MOBILESoft}.
The only difference is code features, and ORLIS uses the fuzzy method signature. 
\textit{Juxtapp}~\cite{Juxtapp13DIMVA} also adopts opcode as the feature and applies feature hashing to generate fingerprints to recognize the clone apps. \textit{LibD}~\cite{LibD2017ICSE,LibD22018TSE} and the library detection system developed by Han et al.~\cite{identifyads2018WPC} both construct the control flow graph (CFG) and extract the opcode from each basic block as the code feature. 
Based on our research, we find that existing detection tools all exploit the hash algorithm to generate the library fingerprints.
\textit{DroidSim}~\cite{DroidSim2014IFIP} uses the component-based control-flow graph (CB-CFG) as the fingerprint and VF2~\cite{Vf22004} a subgraph isomorphism algorithm to identify the clone apps. However, VF2 is not good for large-scale detection. 
Chen et al.~\cite{chen14ICSE} designed a new data structure (3D-CFG) to replace general CFGs and introduced a new algorithm \textit{centroid} that converts the 3D-CFG into a vector to represent a CFG. This method is also adopted by OSSPoLICE~\cite{OSSPOLICE2017CCS}.

\settowidth\rotheadsize{\theadfont{LibExtractorrr}}
\begin{table*}[t]
\centering
\caption{The comparison of existing state-of-the-art TPL detection tools}
\vspace{-1ex}
\scalebox{0.88}{
\begin{tabular}{cccccccccccccc}
\toprule[1.3pt]
\multicolumn{1}{l}{} &  & \rothead{\textbf{LibExtractor}} & \rothead{\textbf{LibRoad}} & \rothead{\textbf{LibID}} & \rothead{\textbf{LibPecker}} & \rothead{\textbf{ORLIS}} & \rothead{\textbf{Han et al.}} & \rothead{\textbf{OSSPOLICE}} & \rothead{\textbf{LibD}} & \rothead{\textbf{LibScout}} & \rothead{\textbf{LibRadar}} & \rothead{\textbf{LibSift}}  & \rothead{\textbf{AdDetect}} \\ 
\midrule[1.3pt]
\Xhline{0.4mm}
\multicolumn{2}{|c|}{\textbf{Year}} & \multicolumn{1}{c|}{2020} & \multicolumn{1}{c|}{2020} & \multicolumn{1}{c|}{2019} & \multicolumn{1}{c|}{2018} & \multicolumn{1}{c|}{2018} & \multicolumn{1}{c|}{2018} & \multicolumn{1}{c|}{2017} & \multicolumn{1}{c|}{19,17} & \multicolumn{1}{c|}{2016} & \multicolumn{1}{c|}{2016} & \multicolumn{1}{c|}{2016} & \multicolumn{1}{c|}{2014} \\ \hline
\multicolumn{2}{|c|}{\textbf{Tool Available}} & \multicolumn{1}{c|}{} & \multicolumn{1}{c|}{} & \multicolumn{1}{c|}{\cmark} & \multicolumn{1}{c|}{\cmark} & \multicolumn{1}{c|}{\cmark} & \multicolumn{1}{c|}{} & \multicolumn{1}{c|}{\cmark} & \multicolumn{1}{c|}{\cmark} & \multicolumn{1}{c|}{\cmark} & \multicolumn{1}{c|}{\cmark} & \multicolumn{1}{c|}{} & \multicolumn{1}{c|}{} \\ \hline
\multicolumn{2}{|c|}{\textbf{Dataset Available}} & \multicolumn{1}{c|}{} & \multicolumn{1}{c|}{} & \multicolumn{1}{c|}{} & \multicolumn{1}{c|}{} & \multicolumn{1}{c|}{\cmark} & \multicolumn{1}{c|}{} & \multicolumn{1}{c|}{} & \multicolumn{1}{c|}{} & \multicolumn{1}{c|}{} & \multicolumn{1}{c|}{} & \multicolumn{1}{c|}{} & \multicolumn{1}{c|}{} \\ \Xhline{0.3mm}

\multicolumn{1}{|c|}{\multirow{6}{*}{\textbf{\begin{tabular}[c]{@{}c@{}}Pre-processing\\ Tool\end{tabular}}}} & \multicolumn{1}{c|}{\textbf{Apktool}} & \multicolumn{1}{c|}{} & \multicolumn{1}{c|}{\cmark} & \multicolumn{1}{c|}{} & \multicolumn{1}{c|}{\cmark} & \multicolumn{1}{c|}{} & \multicolumn{1}{c|}{} & \multicolumn{1}{c|}{} & \multicolumn{1}{c|}{\cmark} & \multicolumn{1}{c|}{} & \multicolumn{1}{c|}{} & \multicolumn{1}{c|}{\cmark} & \multicolumn{1}{c|}{\cmark} \\ \cline{2-14} 
\multicolumn{1}{|c|}{} & \multicolumn{1}{c|}{\textbf{Androguard}} & \multicolumn{1}{c|}{} & \multicolumn{1}{c|}{} & \multicolumn{1}{c|}{\cmark} & \multicolumn{1}{c|}{\cmark} & \multicolumn{1}{c|}{} & \multicolumn{1}{c|}{\cmark} & \multicolumn{1}{c|}{} & \multicolumn{1}{c|}{\cmark} & \multicolumn{1}{c|}{} & \multicolumn{1}{c|}{} & \multicolumn{1}{c|}{} & \multicolumn{1}{c|}{} \\ \cline{2-14} 
\multicolumn{1}{|c|}{} & \multicolumn{1}{c|}{\textbf{Soot}} & \multicolumn{1}{c|}{} & \multicolumn{1}{c|}{} & \multicolumn{1}{c|}{} & \multicolumn{1}{c|}{} & \multicolumn{1}{c|}{\cmark} & \multicolumn{1}{c|}{} & \multicolumn{1}{c|}{\cmark} & \multicolumn{1}{c|}{} & \multicolumn{1}{c|}{\cmark} & \multicolumn{1}{c|}{} & \multicolumn{1}{c|}{} & \multicolumn{1}{c|}{} \\ \cline{2-14} 
\multicolumn{1}{|c|}{} & \multicolumn{1}{c|}{\textbf{baksmali}} & \multicolumn{1}{c|}{\cmark} & \multicolumn{1}{c|}{} & \multicolumn{1}{c|}{} & \multicolumn{1}{c|}{} & \multicolumn{1}{c|}{} & \multicolumn{1}{c|}{} & \multicolumn{1}{c|}{} & \multicolumn{1}{c|}{} & \multicolumn{1}{c|}{} & \multicolumn{1}{c|}{} & \multicolumn{1}{c|}{} & \multicolumn{1}{c|}{} \\ \cline{2-14} 
\multicolumn{1}{|c|}{} & \multicolumn{1}{c|}{\textbf{dex2jar}} & \multicolumn{1}{c|}{} & \multicolumn{1}{c|}{} & \multicolumn{1}{c|}{\cmark} & \multicolumn{1}{c|}{} & \multicolumn{1}{c|}{} & \multicolumn{1}{c|}{} & \multicolumn{1}{c|}{} & \multicolumn{1}{c|}{} & \multicolumn{1}{c|}{} & \multicolumn{1}{c|}{} & \multicolumn{1}{c|}{} & \multicolumn{1}{c|}{} \\ \cline{2-14} 
\multicolumn{1}{|c|}{} & \multicolumn{1}{c|}{\textbf{LIBDEX}} & \multicolumn{1}{c|}{} & \multicolumn{1}{c|}{} & \multicolumn{1}{c|}{} & \multicolumn{1}{c|}{} & \multicolumn{1}{c|}{} & \multicolumn{1}{c|}{} & \multicolumn{1}{c|}{} & \multicolumn{1}{c|}{} & \multicolumn{1}{c|}{} & \multicolumn{1}{c|}{\cmark} & \multicolumn{1}{c|}{} & \multicolumn{1}{c|}{} \\ \Xhline{0.35mm}
\multicolumn{1}{|c|}{\multirow{4}{*}{\textbf{\begin{tabular}[c]{@{}c@{}}Module \\ Decoupling\\ Feature\end{tabular}}}} & \multicolumn{1}{c|}{\textbf{PHS,PN}} & \multicolumn{1}{c|}{} & \multicolumn{1}{c|}{\cmark} & \multicolumn{1}{c|}{\cmark} & \multicolumn{1}{c|}{\cmark} & \multicolumn{1}{c|}{} & \multicolumn{1}{c|}{} & \multicolumn{1}{c|}{\cmark} & \multicolumn{1}{c|}{} & \multicolumn{1}{c|}{\cmark} & \multicolumn{1}{c|}{\cmark} & \multicolumn{1}{c|}{} & \multicolumn{1}{c|}{} \\ \cline{2-14} 
\multicolumn{1}{|c|}{} & \multicolumn{1}{c|}{\textbf{Homogeny Graph}} & \multicolumn{1}{c|}{} & \multicolumn{1}{c|}{} & \multicolumn{1}{c|}{} & \multicolumn{1}{c|}{} & \multicolumn{1}{c|}{} & \multicolumn{1}{c|}{} & \multicolumn{1}{c|}{} & \multicolumn{1}{c|}{\cmark} & \multicolumn{1}{c|}{} & \multicolumn{1}{c|}{} & \multicolumn{1}{c|}{} & \multicolumn{1}{c|}{} \\ \cline{2-14} 
\multicolumn{1}{|c|}{} & \multicolumn{1}{c|}{\textbf{PDG}} & \multicolumn{1}{c|}{} & \multicolumn{1}{c|}{} & \multicolumn{1}{c|}{} & \multicolumn{1}{c|}{} & \multicolumn{1}{c|}{} & \multicolumn{1}{c|}{} & \multicolumn{1}{c|}{} & \multicolumn{1}{c|}{} & \multicolumn{1}{c|}{} & \multicolumn{1}{c|}{} & \multicolumn{1}{c|}{\cmark} & \multicolumn{1}{c|}{\cmark} \\ \cline{2-14} 
\multicolumn{1}{|c|}{} & \multicolumn{1}{c|}{\textbf{Class Dependency}} & \multicolumn{1}{c|}{\cmark} & \multicolumn{1}{c|}{} & \multicolumn{1}{c|}{} & \multicolumn{1}{c|}{} & \multicolumn{1}{c|}{\cmark} & \multicolumn{1}{c|}{} & \multicolumn{1}{c|}{} & \multicolumn{1}{c|}{} & \multicolumn{1}{c|}{} & \multicolumn{1}{c|}{} & \multicolumn{1}{c|}{} & \multicolumn{1}{c|}{} \\ \Xhline{0.3mm}
\multicolumn{1}{|c|}{\multirow{2}{*}{\textbf{\begin{tabular}[c]{@{}c@{}}Decoupling\\ Method\end{tabular}}}} & \multicolumn{1}{c|}{\textbf{BIP}} & \multicolumn{1}{c|}{} & \multicolumn{1}{c|}{} & \multicolumn{1}{c|}{\cmark} & \multicolumn{1}{c|}{} & \multicolumn{1}{c|}{} & \multicolumn{1}{c|}{} & \multicolumn{1}{c|}{} & \multicolumn{1}{c|}{} & \multicolumn{1}{c|}{} & \multicolumn{1}{c|}{} & \multicolumn{1}{c|}{} & \multicolumn{1}{c|}{} \\ \cline{2-14} 
\multicolumn{1}{|c|}{} & \multicolumn{1}{c|}{\textbf{HAC}} & \multicolumn{1}{c|}{} & \multicolumn{1}{c|}{} & \multicolumn{1}{c|}{} & \multicolumn{1}{c|}{} & \multicolumn{1}{c|}{} & \multicolumn{1}{c|}{} & \multicolumn{1}{c|}{} & \multicolumn{1}{c|}{} & \multicolumn{1}{c|}{} & \multicolumn{1}{c|}{} & \multicolumn{1}{c|}{\cmark} & \multicolumn{1}{c|}{\cmark} \\ \Xhline{0.3mm}
\multicolumn{1}{|c|}{\multirow{7}{*}{\textbf{\begin{tabular}[c]{@{}c@{}}Extracted \\ Features\end{tabular}}}} & \multicolumn{1}{c|}{\textbf{Fuzzy Method Sig.}} & \multicolumn{1}{c|}{\textbf{}} & \multicolumn{1}{c|}{\textbf{\cmark}} & \multicolumn{1}{c|}{\textbf{\cmark}} & \multicolumn{1}{c|}{\textbf{}} & \multicolumn{1}{c|}{\textbf{\cmark}} & \multicolumn{1}{c|}{\textbf{}} & \multicolumn{1}{c|}{\textbf{\cmark}} & \multicolumn{1}{c|}{\textbf{}} & \multicolumn{1}{c|}{\textbf{\cmark}} & \multicolumn{1}{c|}{\textbf{}} & \multicolumn{1}{c|}{\textbf{}} & \multicolumn{1}{c|}{\textbf{}} \\ \cline{2-14} 
\multicolumn{1}{|c|}{} & \multicolumn{1}{c|}{\textbf{CFG}} & \multicolumn{1}{c|}{\textbf{}} & \multicolumn{1}{c|}{\textbf{}} & \multicolumn{1}{c|}{\textbf{\cmark}} & \multicolumn{1}{c|}{\textbf{}} & \multicolumn{1}{c|}{\textbf{}} & \multicolumn{1}{c|}{\textbf{\cmark}} & \multicolumn{1}{c|}{\textbf{}} & \multicolumn{1}{c|}{\textbf{\cmark}} & \multicolumn{1}{c|}{\textbf{}} & \multicolumn{1}{c|}{\textbf{}} & \multicolumn{1}{c|}{\textbf{}} & \multicolumn{1}{c|}{\textbf{}} \\ \cline{2-14} 
\multicolumn{1}{|c|}{} & \multicolumn{1}{c|}{\textbf{APIs}} & \multicolumn{1}{c|}{\textbf{}} & \multicolumn{1}{c|}{\textbf{}} & \multicolumn{1}{c|}{\textbf{}} & \multicolumn{1}{c|}{\textbf{}} & \multicolumn{1}{c|}{\textbf{}} & \multicolumn{1}{c|}{\textbf{}} & \multicolumn{1}{c|}{\textbf{}} & \multicolumn{1}{c|}{\textbf{}} & \multicolumn{1}{c|}{\textbf{}} & \multicolumn{1}{c|}{\textbf{\cmark}} & \multicolumn{1}{c|}{\textbf{}} & \multicolumn{1}{c|}{\textbf{\cmark}} \\ \cline{2-14} 
\multicolumn{1}{|c|}{} & \multicolumn{1}{c|}{\textbf{Class Dependency}} & \multicolumn{1}{c|}{\textbf{\cmark}} & \multicolumn{1}{c|}{\textbf{}} & \multicolumn{1}{c|}{\textbf{\cmark}} & \multicolumn{1}{c|}{\textbf{\cmark}} & \multicolumn{1}{c|}{\textbf{}} & \multicolumn{1}{c|}{\textbf{}} & \multicolumn{1}{c|}{\textbf{}} & \multicolumn{1}{c|}{\textbf{}} & \multicolumn{1}{c|}{\textbf{}} & \multicolumn{1}{c|}{\textbf{}} & \multicolumn{1}{c|}{\textbf{}} & \multicolumn{1}{c|}{\textbf{}} \\ \cline{2-14} 
\multicolumn{1}{|c|}{} & \multicolumn{1}{c|}{\textbf{CFG Centroid}} & \multicolumn{1}{c|}{\textbf{}} & \multicolumn{1}{c|}{\textbf{}} & \multicolumn{1}{c|}{\textbf{}} & \multicolumn{1}{c|}{\textbf{}} & \multicolumn{1}{c|}{\textbf{}} & \multicolumn{1}{c|}{\textbf{}} & \multicolumn{1}{c|}{\textbf{\cmark}} & \multicolumn{1}{c|}{\textbf{}} & \multicolumn{1}{c|}{\textbf{}} & \multicolumn{1}{c|}{\textbf{}} & \multicolumn{1}{c|}{\textbf{}} & \multicolumn{1}{c|}{\textbf{}} \\ \cline{2-14} 
\multicolumn{1}{|c|}{} & \multicolumn{1}{c|}{\textbf{Permission,component,UI}} & \multicolumn{1}{c|}{\textbf{}} & \multicolumn{1}{c|}{\textbf{}} & \multicolumn{1}{c|}{\textbf{}} & \multicolumn{1}{c|}{\textbf{}} & \multicolumn{1}{c|}{\textbf{}} & \multicolumn{1}{c|}{\textbf{}} & \multicolumn{1}{c|}{\textbf{}} & \multicolumn{1}{c|}{\textbf{}} & \multicolumn{1}{c|}{\textbf{}} & \multicolumn{1}{c|}{\textbf{}} & \multicolumn{1}{c|}{\textbf{}} & \multicolumn{1}{c|}{\textbf{\cmark}} \\ \cline{2-14} 
\multicolumn{1}{|c|}{} & \multicolumn{1}{c|}{\textbf{Strings}} & \multicolumn{1}{c|}{\textbf{}} & \multicolumn{1}{c|}{\textbf{}} & \multicolumn{1}{c|}{\textbf{}} & \multicolumn{1}{c|}{\textbf{}} & \multicolumn{1}{c|}{\textbf{}} & \multicolumn{1}{c|}{\textbf{}} & \multicolumn{1}{c|}{\textbf{\cmark}} & \multicolumn{1}{c|}{\textbf{}} & \multicolumn{1}{c|}{\textbf{}} & \multicolumn{1}{c|}{\textbf{}} & \multicolumn{1}{c|}{\textbf{}} & \multicolumn{1}{c|}{\textbf{\cmark}} \\ \Xhline{0.3mm}
\multicolumn{1}{|c|}{\multirow{3}{*}{\textbf{\begin{tabular}[c]{@{}c@{}}Referred TPL\\ Construction\\ Method\end{tabular}}}} & \multicolumn{1}{c|}{\textbf{classification}} & \multicolumn{1}{c|}{\textbf{}} & \multicolumn{1}{c|}{\textbf{}} & \multicolumn{1}{c|}{\textbf{}} & \multicolumn{1}{c|}{\textbf{}} & \multicolumn{1}{c|}{\textbf{}} & \multicolumn{1}{c|}{\textbf{}} & \multicolumn{1}{c|}{\textbf{}} & \multicolumn{1}{c|}{\textbf{}} & \multicolumn{1}{c|}{\textbf{}} & \multicolumn{1}{c|}{\textbf{}} & \multicolumn{1}{c|}{\textbf{}} & \multicolumn{1}{c|}{\textbf{\cmark}} \\ \cline{2-14} 
\multicolumn{1}{|c|}{} & \multicolumn{1}{c|}{\textbf{clustering}} & \multicolumn{1}{c|}{\textbf{\cmark}} & \multicolumn{1}{c|}{\textbf{}} & \multicolumn{1}{c|}{\textbf{}} & \multicolumn{1}{c|}{\textbf{}} & \multicolumn{1}{c|}{\textbf{}} & \multicolumn{1}{c|}{\textbf{}} & \multicolumn{1}{c|}{\textbf{}} & \multicolumn{1}{c|}{\textbf{\cmark}} & \multicolumn{1}{c|}{\textbf{}} & \multicolumn{1}{c|}{\textbf{\cmark}} & \multicolumn{1}{c|}{\textbf{}} & \multicolumn{1}{c|}{\textbf{}} \\ \cline{2-14} 
\multicolumn{1}{|c|}{} & \multicolumn{1}{c|}{\textbf{similarity comparison}} & \multicolumn{1}{c|}{\textbf{}} & \multicolumn{1}{c|}{\textbf{\cmark}} & \multicolumn{1}{c|}{\textbf{\cmark}} & \multicolumn{1}{c|}{\textbf{\cmark}} & \multicolumn{1}{c|}{\textbf{\cmark}} & \multicolumn{1}{c|}{\textbf{\cmark}} & \multicolumn{1}{c|}{\textbf{\cmark}} & \multicolumn{1}{c|}{\textbf{}} & \multicolumn{1}{c|}{\textbf{\cmark}} & \multicolumn{1}{c|}{\textbf{}} & \multicolumn{1}{c|}{\textbf{-}} & \multicolumn{1}{c|}{\textbf{}} \\ \Xhline{0.3mm}
\multicolumn{1}{|c|}{\multirow{5}{*}{\textbf{\begin{tabular}[c]{@{}c@{}}Comparison\\ Method\end{tabular}}}} & \multicolumn{1}{c|}{\textbf{Similarity Comparison}} & \multicolumn{1}{c|}{\textbf{\cmark}} & \multicolumn{1}{c|}{} & \multicolumn{1}{c|}{} & \multicolumn{1}{c|}{} & \multicolumn{1}{c|}{} & \multicolumn{1}{c|}{\textbf{\cmark}} & \multicolumn{1}{c|}{} & \multicolumn{1}{c|}{\textbf{\cmark}} & \multicolumn{1}{c|}{} & \multicolumn{1}{c|}{\textbf{\cmark}} & \multicolumn{1}{c|}{-} & \multicolumn{1}{c|}{-} \\ \cline{2-14} 
\multicolumn{1}{|c|}{} & \multicolumn{1}{c|}{\textbf{\begin{tabular}[c]{@{}c@{}}Fuzzy CLass Match\\ (Adaptive Match)\end{tabular}}} & \multicolumn{1}{c|}{} & \multicolumn{1}{c|}{\textbf{\cmark}} & \multicolumn{1}{c|}{} & \multicolumn{1}{c|}{\textbf{\cmark}} & \multicolumn{1}{c|}{} & \multicolumn{1}{c|}{} & \multicolumn{1}{c|}{} & \multicolumn{1}{c|}{} & \multicolumn{1}{c|}{} & \multicolumn{1}{c|}{} & \multicolumn{1}{c|}{-} & \multicolumn{1}{c|}{-} \\ \cline{2-14} 
\multicolumn{1}{|c|}{} & \multicolumn{1}{c|}{\textbf{Fuzzy Hash}} & \multicolumn{1}{c|}{} & \multicolumn{1}{c|}{} & \multicolumn{1}{c|}{} & \multicolumn{1}{c|}{} & \multicolumn{1}{c|}{\textbf{\cmark}} & \multicolumn{1}{c|}{} & \multicolumn{1}{c|}{} & \multicolumn{1}{c|}{} & \multicolumn{1}{c|}{} & \multicolumn{1}{c|}{} & \multicolumn{1}{c|}{-} & \multicolumn{1}{c|}{-} \\ \cline{2-14} 
\multicolumn{1}{|c|}{} & \multicolumn{1}{c|}{\textbf{Hierarchical Indexing}} & \multicolumn{1}{c|}{} & \multicolumn{1}{c|}{} & \multicolumn{1}{c|}{} & \multicolumn{1}{c|}{} & \multicolumn{1}{c|}{} & \multicolumn{1}{c|}{} & \multicolumn{1}{c|}{\textbf{\cmark}} & \multicolumn{1}{c|}{} & \multicolumn{1}{c|}{} & \multicolumn{1}{c|}{} & \multicolumn{1}{c|}{-} & \multicolumn{1}{c|}{-} \\ \cline{2-14} 
\multicolumn{1}{|c|}{} & \multicolumn{1}{c|}{\textbf{LSH}} & \multicolumn{1}{c|}{} & \multicolumn{1}{c|}{} & \multicolumn{1}{c|}{\textbf{\cmark}} & \multicolumn{1}{c|}{} & \multicolumn{1}{c|}{} & \multicolumn{1}{c|}{} & \multicolumn{1}{c|}{} & \multicolumn{1}{c|}{} & \multicolumn{1}{c|}{} & \multicolumn{1}{c|}{} & \multicolumn{1}{c|}{-} & \multicolumn{1}{c|}{-} \\ \Xhline{0.3mm}
\multicolumn{1}{|c|}{\multirow{5}{*}{\textbf{\begin{tabular}[c]{@{}c@{}}Identification\\ Granularity\end{tabular}}}} & \multicolumn{1}{c|}{\textbf{Class-level}} & \multicolumn{1}{c|}{} & \multicolumn{1}{c|}{} & \multicolumn{1}{c|}{} & \multicolumn{1}{c|}{} & \multicolumn{1}{c|}{\textbf{\cmark}} & \multicolumn{1}{c|}{} & \multicolumn{1}{c|}{} & \multicolumn{1}{c|}{} & \multicolumn{1}{c|}{} & \multicolumn{1}{c|}{} & \multicolumn{1}{c|}{} & \multicolumn{1}{c|}{} \\ \cline{2-14} 
\multicolumn{1}{|c|}{} & \multicolumn{1}{c|}{\textbf{Ad Library}} & \multicolumn{1}{c|}{} & \multicolumn{1}{c|}{} & \multicolumn{1}{c|}{} & \multicolumn{1}{c|}{} & \multicolumn{1}{c|}{} & \multicolumn{1}{c|}{} & \multicolumn{1}{c|}{} & \multicolumn{1}{c|}{} & \multicolumn{1}{c|}{} & \multicolumn{1}{c|}{} & \multicolumn{1}{c|}{} & \multicolumn{1}{c|}{\textbf{\cmark}} \\ \cline{2-14} 
\multicolumn{1}{|c|}{} & \multicolumn{1}{c|}{\textbf{pacakge-level}} & \multicolumn{1}{c|}{\textbf{\cmark}} & \multicolumn{1}{c|}{\textbf{}} & \multicolumn{1}{c|}{} & \multicolumn{1}{c|}{\textbf{}} & \multicolumn{1}{c|}{} & \multicolumn{1}{c|}{} & \multicolumn{1}{c|}{} & \multicolumn{1}{c|}{\textbf{\cmark}} & \multicolumn{1}{c|}{} & \multicolumn{1}{c|}{\textbf{\cmark}} & \multicolumn{1}{c|}{} & \multicolumn{1}{c|}{} \\ \cline{2-14} 
\multicolumn{1}{|c|}{} & \multicolumn{1}{c|}{\textbf{Library-level}} & \multicolumn{1}{c|}{\textbf{\cmark}} & \multicolumn{1}{c|}{\textbf{\cmark}} & \multicolumn{1}{c|}{\textbf{}} & \multicolumn{1}{c|}{\textbf{\cmark}} & \multicolumn{1}{c|}{} & \multicolumn{1}{c|}{} & \multicolumn{1}{c|}{\textbf{}} & \multicolumn{1}{c|}{\textbf{\cmark}} & \multicolumn{1}{c|}{\textbf{}} & \multicolumn{1}{c|}{\textbf{\cmark}} & \multicolumn{1}{c|}{} & \multicolumn{1}{c|}{} \\ \cline{2-14} 
\multicolumn{1}{|c|}{} & \multicolumn{1}{c|}{\textbf{Version-level}} & \multicolumn{1}{c|}{} & \multicolumn{1}{c|}{} & \multicolumn{1}{c|}{\textbf{\cmark}} & \multicolumn{1}{c|}{} & \multicolumn{1}{c|}{} & \multicolumn{1}{c|}{} & \multicolumn{1}{c|}{\textbf{\cmark}} & \multicolumn{1}{c|}{} & \multicolumn{1}{c|}{\textbf{\cmark}} & \multicolumn{1}{c|}{} & \multicolumn{1}{c|}{} & \multicolumn{1}{c|}{} \\ \Xhline{0.3mm}
\end{tabular}
}
	\begin{center}
		\textit{ PN: package name; PHS: package hierarchy structure; PDG: package dependency graph;  \cmark: represents the use of a tool, a feature or a technique \\ LSH: Locality-Sensitive Hashing: HAC: Hierarchy Agglomerative Clustering; BIP: Binary Integer Programming models '-': not applicable. \\}
	\end{center}
	\label{tbl:TPL_tech_CMP}
\vspace{-3ex}
\end{table*}

\section{Tool Comparison}
\label{sec:comparison}

\subsection{Overview of TPL Detection Tools}
\label{sec:overview}

To investigate existing TPL detection techniques, we first follow a well-defined Systematic Literature Review (SLR) methodology~\cite{snowballing2014,SLR2007} to find related research in this area. We search the candidate papers from four digital databases: ACM Digital Library, IEEE Xplore, SpringerLink, and ScienceDirect by using keywords (e.g., third-party library) search. Besides, we also search the relevant publications from top-tier venues on software engineering~\cite{sevenue}, security~\cite{secvenue}. The two links are the list of considered conferences and journals.
We do not consider posters~\cite{MOBSCANNER2017ICSE-C} or short papers that provide a preliminary idea. Furthermore, we do not consider that papers focus on the native library (developed by C/C++) detection since the detection techniques are quite different. Thus, we exclude papers such as LibDX~\cite{LibDX2020Saner} and BAT~\cite{BAT2011MSR}. 
Finally, we get 12 relevant detection tools. $LibD^{2}$~\cite{LibD22018TSE} is an extension of LibD~\cite{LibD2017ICSE}, therefore, we discuss them together.
We introduce and compare these TPL detection techniques, with details shown in Table~\ref{tbl:TPL_tech_CMP}.

{Strictly speaking, various TPL detection techniques began to appear one after another since 2016 because AdDetect~\cite{AdDetect2014ISSNIP} and LibSift~\cite{LibSift2016soh} cannot identify specific TPLs.
AdDetect just picks ad libraries from the host app  by using a binary classification algorithm (i.e., SVM~\cite{SVM}). LibSift only can separate the TPLs from host app by implementing a module decoupling algorithm.}
{According to the TABLE~\ref{tbl:TPL_tech_CMP}, we can find that more than half (7/12) of the tools are publicly available, but seldom of them published their dataset except ORLIS published a part of dataset.
{ORLIS provides two datasets; the first one is hash values of the closed-source apps and corresponding names of TPLs; the second one is a set of apps and corresponding obfuscated apps that are obfuscated by Proguard, Allatori and DashO, respectively. However, We cannot directly use the first dataset to evaluate existing tools because ORLIS did not offer the mapping information between apps and TPLs and only give the hash value of apps and the names of TPLs. It is difficult to ensure the correct TPLs and down them from repositories if we only know the names of TPLs, because even the different TPLs may have the same name.} 
LibScout published the library profiles (a.k.a, library signature)~\cite{libscout-profile} of their experimental dataset but provided a script \textsc{library-scraper} that can help users download original library SDKs from maven repository~\cite{maven}. We also cannot directly use this dataset due to lack of the apps.}
{For the remaining subsections, we compare TPL detection tools based on the detection process.}

\subsection{Preprocessing Comparison} 
In the preprocessing stage, we can find from Table~\ref{tbl:TPL_tech_CMP} that Apktool~\cite{apktool} is the most frequently-used tool (5/12). Androguard~\cite{Androguard} can be used to generate the class dependency relationship; both Androguard and Soot~\cite{soot} can be used to construct the control flow graph (CFG). 

{
Java exploits the package to organize class files into different namespaces, which means TPLs usually have their independent package structure.
The package structure is a tree hierarchy structure. The lower levels in this tree structure are considered as sub-packages. Each node on the directory tree can be the java class file or the sub-directories, which provides a convenience to integrate library code into the host apps in a static link method. Based on the package hierarchy structure information, we can roughly estimate whether an app includes TPLs or not. Thus, researchers usually choose apktool~\cite{apktool} and Androguard~\cite{Androguard} as the reverse-engineering tools to decompile apks because they can help reproduce the package structure of TPLs.
}

\vspace{-2mm}

\subsection{Library Instance Construction Comparison}
As shown in Table~\ref{tbl:TPL_tech_CMP} on library instance construction, apart from the package name (PN), another {four} features are used to identify the boundaries of TPLs: (1) \textit{package hierarchy structure (PHS)}, (2) \textit{homogeny graph}, (3) \textit{package dependency graph (PDG)} and (4) \textit{class dependency}.

\noindent $\bullet$ \textbf{Package Hierarchy Structure (PHS)} is a tree, which can be treated as a graph where each node indicates a package, a sub-package or a file, and each edge indicates the inclusion relations between two nodes. {We can find about 42\% of tools (i.e., LibRoad, LibID, LibPecker, \cite{identifyads2018WPC}, LibRadar) use each independent directory tree as a library instance candidate.} Using PHS as the feature to construct the library candidate is a very straightforward idea. In general, an independent TPL usually has an independent PHS; thus, using package structure as the split feature does make sense. However, when various TPLs are imported into an app, things will be different. We have given some motivating examples in Section~\textbf{\ref{Sec:challenge}}. {For example, different TPLs may share the same package tree. A TPL may include several package trees due to TPL dependencies. The code obfuscation can change the package hierarchy structures. The package structures of different versions of the same TPLs could be mutant. Using package structure actually is not very accurate and is difficult to find the correct boundaries of different TPLs, which may compromise the detection performance.}

\noindent $\bullet$ \textbf{Homogeny Graph}, which indicates the parent or sibling relations between two nodes in a graph. The concept of the homogeny graph is proposed by LibD~\cite{LibD2017ICSE}. In a homogeny graph, each node indicates a package or a class file, and the edge means two packages have inclusion or inheritance relations.
{This method first constructs the homogeny graph by integrating nodes with inclusion and inheritance relationships, and then uses class dependencies to construct the complete homogeny graph. We can see that the homogeny graph considers TPL dependencies.}

\noindent $\bullet$ \textbf{Package Dependency Graph (PDG)} includes the class dependency and intra-package homogeny relation~\cite{LibSift2016soh} at the same time. The class dependency includes member field reference relation, method invocation relation, class interface and inheritance relation. Both LibSift and AdDetect adopt this method to conduct the module decoupling.
{Choosing the PDGs as the split features are usually need to depend on the hierarchy agglomerative clustering (HAC) algorithm to conduct the module decoupling.
The PDG also can be treated as a tree, and different dependency relations will be set different weights based on intimacy. The weight setting can refer to original papers~\cite{AdDetect2014ISSNIP,LibSift2016soh}. HAC can cut the non-primary module into different parts based on the intimacy weight.}


\noindent $\bullet$ \textbf{Class Dependency} usually includes the class inheritance, class interface, method invocation, and field reference relationship. Note that not all tools include all of these class dependencies; different tools may have some subtle differences in choosing the aforementioned class dependencies. PDGs also include class dependency. Thus, LibExtractor, ORLIS, LibSift, and AdDetect all use the class dependency in module decoupling. 

\vspace{-1ex}
\subsection{Feature Extraction Comparison}

We compare the feature extraction process of existing TPL detection methods from two aspects: signature representation and feature generation method. 


\noindent $\bullet$ \textbf{Signature Representation.}
As shown in TABLE~\ref{tbl:TPL_tech_CMP}, {we can see that many tools overlap when selecting features of TPLs. The frequently-used features are class dependency, CFG, and fuzzy method signature.} 

{Three tools (i.e., LibExtractor, LibID, and LibPecker) leverage class dependency relations as features but they adopt different hash algorithms to generate signatures.
LibID and LibPecker adopt class inheritance dependency, field dependency, and method invocation dependency. Apart from above mentioned features, LibExtractor adds the class interface relationship in code features.}

{Four tools (i.e., LibRoad, ORLIS, LibScout, and OSSPoLICE) choose the method signature as the TPL features but with different generation methods.
ORLIS, LibScout, and OSSPoLICS select the fuzzy method signature as the feature. The fuzzy method signature is a variant of the method signature, which means using a placeholder, such as X, to replace the identifier that can be easily changed by the obfuscators in the original method signature. These replaced variables are usually developer-defined variables and types; the system-defined hard-code types remain unchanged during obfuscation. ORLIS and LibScout adopt the same code feature but different code feature generation methods.
OSSPoLICE extracts two-stage code features to identify specific TPL versions. OSSPoLICE first uses string constants and fuzzy method signatures to construct the library-level signature. And then, it adopts the CFG centroid to identify specific versions in the second stage. The CFG centroid is a three-dimensional vector, which can be obtained by calculating the centroid of a CFG. In a CFG, each node is called a basic block and can be indicated by a three-dimensional data point, representing the node index, outgoing degree, and loop depth, respectively. LibRoad extracts all method signatures and field signatures in each class and then connects them into a string as the class feature.}

LibD and Han et al.~\cite{identifyads2018WPC} use opcode from control flow graph (CFG) blocks as features and use hash methods to generate the opcode. The only difference is that besides the opcode, Han et al. adopt \textit{Method Type Tag}~\cite{identifyads2018WPC} as well.

LibRadar exploits the Android APIs, the total number of Android APIs and API types to construct the feature vector. These features are not easily modified during obfuscation and can indicate the low-level behaviors of TPLs.

AdDetect extracts app component usages information, device identifiers and users' profile, Android permissions, as well as Android APIs to represents ad library features.

\noindent $\bullet$ \textbf{Feature Generation Method.}
Based on Table~\ref{tbl:TPL_tech_CMP}, we can find most systems adopt hash values to represent features because hash values can dramatically reduce the storage space and search consumption.

{LibExtractor designs an algorithm that contains four-round calculations for each class in a TPL candidate. For the first round, LibExtractor extracts the relative path from this class to each of its dependency classes. LibExtractor also uses a placeholder A to replace all package names in a relative path (``A/A/..''), to be resilient package name obfuscation. These relative paths of each objective class are sorted by their length and then concatenated into a string. Hash this string to get a class signature in the first round. For a class, the class name and their dependency class signatures ( hash values that get in the first round) are concatenated into a new string and hash again as the second round features. For each class, LibExtractor extracts the inheritance and interface classes, direct/virtual methods, instance/static fields and then concatenates them into a string. Hash this string as the third-round feature. The third-round feature connects with the name of this objective class into a string, hash this string as the final features. This method is resilient to renaming obfuscation.
LibRoad generates the final signature by using the MD5 hash function. LibRadar uses feature hashing to generate package-level signatures.}

{LibD and Han et al.~\cite{identifyads2018WPC} extract the opcode sequences from each block of CFGs and calculate the hash of the opcode sequences for each block. In CFGs, each node represents a block and then they choose the depth-first order algorithm to traverse the CFGs and get the method signatures. The feature values of all methods are concatenated in non-decreasing order. The method signatures are hashed again as the class signatures. Finally, all of the class signatures are ordered in non-decreasing order, and the class signatures are hashed again as the library instance fingerprint.}

{ORLIS first uses one feature hash algorithm {(sdhash)~\cite{sdhash}} to hash the fuzzy method signature to represent the library-level signatures and then applies the ssdeep hash algorithm to generate the class-level features.}

LibScout exploits Merkle Tree to generate the TPL features; it uses the bottom-up order to generate the code feature. LibScout adopts the 128-bit MD5 hash to generate method-level features. Similar to LibD, method-level features are sorted and concatenated to re-hash and get the class-level features. Class-level features are sorted and concatenated to the package-level features.

{OSSPoLICE leverages the method of LibScout to generate library-level code features. Based on the library-level signatures, it can get potential TPL candidates. Then, it generates the CFG centroid collection for each TPL candidate. By comparing the similarity of the centroid collection, it can ensure the used TPL versions.}

{Note that LibSift does not identify specific TPLs but split independent TPL candidates out. 
AdDetect employs static analysis to extract the code feature represented as vectors.}

\vspace{-1ex}
\subsection{Library Identification Comparison}

{We introduce library identification from three aspects: referred TPL construction, comparison method and identification granularity.}

\noindent $\bullet$\textbf{Referred TPL Construction.} Based on the strategies of how to generate the referred TPL database, existing tools can be divided into three categories: 1) similarity comparison-based method, 2) clustering-based method, and 3) classification-based method.

{For similarity comparison-based method, it requires developers to collect the TPL files in advance to build the pre-defined database. By using the same feature extraction method on tested apps and the original TPLs to generate the code features. The similarity comparison-based tool compare the features of candidates from host apps and TPLs in  the pre-defined database to confirm  the in-app TPLs.}

The clustering-based method does not require collecting ground-truth TPLs to build a pre-defined database in advance. However, the tool needs to collect apps as input to cluster the potential TPLs. Researchers will manually verify the clustering results and these verified TPL will be treated as the referred TPLs. Therefore, clustering-based tools usually require a substantial number of apps as input to guarantee to generate enough TPL features for feature database.
Table~\ref{tbl:TPL_tech_CMP} shows that LibExtractor, LibD and LibRadar choose the clustering-based method to identify TPLs; and they collect 217,027, 1,427,395 and more than 1 million apps respectively to generate the TPL feature database.
{This method can efficiently find commonly-used in-app TPLs. Nevertheless, it may fail to identify niche and emerging TPLs.}


Classification-based method is only use by AdDetect. It employs SVM algorithm to classify the ad/non-ad libraries.

\noindent$\bullet$\textbf{Identification granularity.} {In this stage, the comparison features have three different granularities: i) the fine-grained features at the class level; ii) the coarse-grained features at the package level; and iii) the hybrid features including the package and class level code features.}

{We can find that all clustering-based tools (i.e., LibExtractor, LibD, and LibRadar) leverage the package-level features.
The similarity comparison-based tools adopt various feature granularity.
LibRoad, OSSPoLICE and LibScout use both package-level and class-level features at comparison stage. For LibRoad, it considers two situations, the package name of TPLs are obfuscated or not.
If the name is obfuscated, the package structure of in-app TPLs may be different from the original TPLs. LibRoad directly compares the class-level features. 
If the package name of TPL is not obfuscated, the in-app TPLs can keep the complete package hierarchy structure. Thus, it uses package hierarchy structure as one of the features and compare the code features at package-level, which can improve the detection efficiency.
OSSPoLICE first adopts the method of LibScout to identify the TPL, and use the CFG centroid method to identify the specific version.
LibScout first uses the package-level granularity to search the similar code features in database. If it cannot find the match pairs, it will compare the code features at class-level.}

\section{Empirical Study Design}
\label{sec: empirical study}

We attempt to thoroughly compare the state-of-the-art TPL detection tools from the practice perspective.
Thus, we design an empirical study by using the following {six} criteria:

\noindent\textbf{C1: Accuracy of Module Decoupling.}
{Based on our previous analysis, we know that existing tools use different module decoupling techniques to construct in-app TPL candidate instances. We have presented the challenges of TPL detection due to the complex package structure. Besides, these challenging cases are very common in real-world apps. Correctly dividing TPLs' boundaries is a prerequisite to ensure that TPL detection tools can achieve a better detection rate. Thus, it is necessary to design an experiment to verify the TPL instance construction accuracy of state-of-the-art tools.}


\noindent\textbf{C2: Effectiveness.} 
To better understand the effectiveness, we compare the existing tools on a unified and unbiased dataset by using three metrics: recall, precision, and F1-Score~\cite{F1_Score}. The equations are as follows:

\begin{equation}
    Precision = \frac{\#\ of\ TPs}{\#\ of\ detected\ TPLs} 
\end{equation}

\begin{equation}
    Recall = \frac{\#\ of\ TPs}{\# of\ TPLs\ in\ ground truth}
\end{equation}

\begin{equation}
    F1-value = \frac{2 \times precision \times recall}{precision + recall}
\end{equation}

{\noindent Specifically, the \textbf{True Positives (TPs)} are the identified TPLs by tools which are also in the ground truth set. The ground truth is the apps and the corresponding used in-app TPLs that we have manually verified. The collected versions of TPLs are used to generate the referred TPL feature database. More specifically, the ground truth for C2 is the second dataset (cf. Sec~\ref{sec:datacollection}).}

\noindent\textbf{C3: Efficiency/Scalability.} Efficiency is an important index to evaluate TPL detection tools.
We compare the detection time of each tool by using the same dataset in C3 and attempt to pick out the tools that are suitable for large-scale detection and can be extended as industrial products.

\noindent\textbf{C4: Capability of version identification.}
Version identification has many significant usages in downstream tasks, such as vulnerable library identification, library updating understanding, outdated TPLs identification. Besides, identifying the specific version is a challenging task because the adjacent versions usually can lead to many false positives.
However, many existing studies lack related evaluation on this aspect. Thus, we try to reveal the version identification capabilities of existing tools in this paper and provide a road map to downstream tasks. We also use the recall, precision and F1-score as the metrics to evaluate them. 

\noindent\textbf{C5: Capability of obfuscation-resilience.} 
Based on a previous study~\cite{Lin2014soups}, more than 50\% TPLs in apps are obfuscated.
Obfuscated TPLs can discount the detection accuracy.  
We thus try to compare the obfuscation-resilience capability of each tool against different obfuscation strategies.
Besides, for the same obfuscation strategy, different {obfuscators ({obfuscation tools})} have various implementation schemes, we also compare the detection ability of existing TPL detection tools against different obfuscators. In this way, we try to figure out the impact of obfuscation on each tools, which tool has the best performance of resiliency to code obfuscation, and what causes this result.

\noindent\textbf{C6: Ease of Use.} Usability of a tool is usually the primary concern for users.
Thus, we reveal the usability of each detection tool by designing a survey to investigate the user experiences and let users rate each tool.

\subsection {Tool Selection}
\label{sec:tools_selection}
Our evaluation only considers the publicly available tools in Table~\ref{tbl:TPL_tech_CMP}, among which LibD is reported containing an error in terms of the hash method by the owner~{\cite{LibD_code}}, we thus excluded {its evaluation result} in this paper. However, we still conduct the experiments for LibD, and the full comparison results can be seen on our website~\cite{LibDetect_web}. OSSPoLICE also reports running errors in the published version~\cite{osspolice} and we contacted the authors many times but failed to get the response, so we do not compare this tool as well.
Eventually, we present the comparison results of five tools (i.e., LibID, LibRadar, LibScout, LibPecker, and ORLIS) in this section. 

\subsection{Data Construction}
\label{sec:datacollection}
The datasets for evaluation need to reflect the challenges in TPL detection and ensure the fair comparisons. Meanwhile the datasets should be able to demonstrate the performance of these tools from different aspects.
Thus, for our empirical study, {four} datasets are collect for different purposes:
(1) Verify the accuracy of library instance construction of different module decoupling techniques when facing possible situations of imported in-app TPLs (for C1). 
(2) Evaluate the effectiveness/efficiency of publicly available tools by detecting the real world apps from Google Play (for C2 \& C3). 
(3) Evaluate the capability of version identification (for C4) by collecting open-source apps and corresponding TPL versions. 
(4) Evaluate the obfuscation-resilient capability of publicly available tools by detecting TPLs in open-source apps with/without obfuscation (for C5).

{We need to build four different datasets because one dataset  cannot meet all the evaluation requirements.
More specifically, for the first dataset, it is difficult to find one or several apps with TPLs cover the various package structures at the same time. Even though we know that a TPL includes TPL dependencies, we still cannot know which apps include this TPL. Thus, we need to build such a dataset by ourselves for some special cases. For example, import some TPLs for host apps. 
For the second dataset, we want to use the commercial apps from Google Play because it is more valuable by using apps from the real world to show the performance of each tool.
For the third dataset, we cannot directly use the aforementioned dataset, because it is difficult to know the specific in-app library versions in commercial apps while we can easily obtain the versions of the used in-app TPLs in open-source apps by checking the config file (e.g., ``build.gradle''). Besides, TPLs in open-source apps and closed-source apps also could be different. For commercial apps published in app markets, app developers can adopt various obfuscators and select different obfuscation strategies to obfuscate in-app TPLs. However, it is difficult for us to know the aforementioned information. Different TPLs usually have different version files, if we just use one dataset to evaluate these tools, the results may not reliable. To make the experimental more sound, we collect two different datasets without intersection to evaluate their capability of version identification.
For the last dateset, it is also difficult for us to know about the collected apps in the previous datasets 1) whether the apps are obfuscated, 2) what obfuscators the developers used, and 3) what obfuscation techniques are adopted.
Thus, we cannot directly use the above-mentioned datasets to evaluate their obfuscation-resilient capability.}
The reasons are as follows:
\textbf{1) Lacks controlled trials.} To evaluate the obfuscation-resilient capability, we need to collect apps with/without code obfuscation. However, it is impossible for the real-world apps from Google Play to meet this condition {because it is difficult for us to know whether these commercial apps are obfuscated or not for some obfuscation techniques (e.g., control flow randomization, API hiding, dead code removal.)}; \textbf{2) Lacks ground truth for code obfuscation.} The apps from Google Play may be obfuscated by developers, and we cannot know which obfuscator they use to obfuscate apps and which obfuscation techniques are adopted. Thus, we cannot use these datasets to evaluate C5.

{In fact, it is not easy to construct such a benchmark, which requires extensive manual work and cross-validation. App obfuscation is also a time-consuming process. It cost us about two months to collect the benchmark.}

\subsubsection{Dataset for Module Decoupling Techniques }
{To verify the in-app TPL candidate construction accuracy of state-of-the-art module decoupling techniques when facing the complex package structures that we mentioned in Section~\ref{Sec:challenge}, we construct a well-designed dataset including TPL dependencies, different TPLs sharing the same root package, and obfuscated TPLs with package flattening. 
}

{For TPL dependencies, we developed two apps and imported the Okhttp V2.0.0~\cite{okhttp2.0.0} into each app. We choose Okhttp V2.0.0 because it depends on another TPL \textit{Okio}~\cite{okio}.
As we mentioned in Section~\ref{sec:challenge:TPL_dependency}, Java has different package techniques. 
To restore all the real-world scenarios of apps invoking TPLs, we designed two groups of experiments.
One app was imported the TPL Okhttp V2.0.0 with ``skinny'' mode, which means the published TPL without the dependency code, but the app includes the code of Okhttp and okio at the same time.
The other app was imported the Okhttp with its direct dependency \textit{Okio}. To achieve that, we modified the source code of Okhttp V2.0.0 via modifying the config file ``pom.xml'' by adding ``jar-with-dependencies'' and recompiled it to let the published Okhttp including the dependency \textit{Okio}. 
}


{For different TPLs with the same root package, we first crawl open-source apps from F-Droid. By analyzing their config files, we try to select apps that include TPLs with similar package structures. Finally, we decide to use the TPLs with the same root package ``com.android.support''. 
We select this kind of TPLs because we find Android apps usually include several Android support TPLs (for instance, ``support-vX'', ``support-appcompat'', ``support-design'') at the same time. These support TPLs usually have the same root package and sub-packages.
For this evaluation, we collect 25 different TPLs and 31 open-source apps.}

{For TPLs with package flattening, we use the fourth dataset that we have used DashO to obfuscate. In addition, the package flattening allows users to modify the internal hierarchy structure or remove the entire package structure.}


\subsubsection{Dataset for Effectiveness/Efficiency Evaluation.}
\label{sec:dataset_for_effectiveness}
This dataset needs to meet two requirements: 1) providing the mapping information between apks and TPLs; and 2) providing a complete version set of each TPL.

Note that we need to collect the TPLs with their complete versions to ensure fairness when comparing these tools. The reasons are as follows: (1) We can only know the libraries used in a commercial app by referring to some websites, such as AppBrain~\cite{AppBrain}, without knowing the specific library version. 
Even for the same TPL, the code similarity of different versions also varies, ranging from 0\% to 100\%. If an app uses TPLs whose versions are not included in the TPL dataset, it could cause false negatives when the code similarity of two versions is below the defined threshold. Thus, to eliminate the side-effects caused by the incomplete versions of TPLs, we should collect the TPLs with their full versions. 
(2) The libraries' updated ways are diverse. Some TPLs require developers to manually update them while some TPLs support automatic update. Therefore, it is difficult to ensure the specific mapping relations between TPLs' version and commercial apps.
(3) Even if different apps include the same TPL, the used version could be different.
We find that only ORLIS released a dataset to evaluate the capability of obfuscation-resilience. However, the number of in-app TPLs from open-source apps is usually small, and most TPLs are non-obfuscated, which cannot reflect the ability of these tools to handle real-world apps. Besides, they do not provide full versions of TPLs in the dataset, which may bias some tools. 
Therefore, we need to collect both the real-world apps and the full versions of used TPLs.

\noindent{\textbf{App Collection.}} 
We refer to the AppBrain~\cite{AppBrain} to get the app-library mapping information and manually check their relationship to ensure the correctness. Then, based on the mapping information, we download 221 Android apps (the newest versions) that use at least one TPL from Google play. 
We need to clarify a confusing concept here. The tools like LibExtractor, LibRadar, and LibD adopt the clustering-based method to identify in-app TPLs, which require a considerable number of apps (million-level) as input to generate enough TPL signatures. Whereas we collect apps to verify the performance of different tools here, the functions of the two sets of apps are different. Thus, we do not need too many apps here. 

\noindent\textbf{TPL Collection.}
According to the download apps, we can get the information of in-app TPLs. We use \texttt{library-scraper}~\cite{libscraper} to crawl TPL files from Maven Central~\cite{maven}, Jcenter~\cite{Jcenter}, Google's maven repository~\cite{GoogleMvn}, etc.
We filter out TPLs whose full versions are not included in our dataset. Finally, we select 59 unique TPLs and 2,115 corresponding library versions. This dataset containing TPLs and the corresponding apps is used as the ground-truth to evaluate the accuracy and performance of each tool.





\subsubsection{Dataset for Version Identification.}
{Based on our observation, code differences among different TPLs are various.
For example, some TPLs usually update frequently and the code differences among the adjacent versions are minor. In contrast, some TPLs do not frequently update their versions, and the code difference among different versions may be apparent.
We intend to verify not only the capabilities of TPL version detection of these tools, but also the performance of each tool towards TPLs with various code differences.}
To this end, we collect two groups of datasets. For the first group, we collect TPLs with obvious code differences among different versions and the corresponding apps. For the second group, we collect TPLs with minor code differences among different versions and their corresponding apps. The specific data collection process is as follows.

\noindent \textbf{TPL Collection.} {We first crawled open-source apps involving 17 different (all) categories from F-Droid~\cite{F-Droid}. 
We filter the apps without TPLs by referring to the config files (e.g., \textit{build.gradle}).
Based on the config files, we can easily verify the used TPLs in open-source apps. We then downloaded the corresponding TPLs and their complete version files from the maven repository.
We calculated the hash value for each class in a version file and compared the similarity value between adjacent versions.
We also adopt the Jaccard index to calculate the similarity value between two versions and each comparison element is the hash value of each class. When the similarity is less than 0.6 between two versions and the version number accounts for  50\% of the total number of all versions of a TPL, we select these TPLs as the first group.
When the similarity is greater than 0.8 and this number account for 70\% of the total number of all versions of a TPL, we select these TPLs as the second group. The threshold selection is based on the collected dataset.
The collected TPLs are used to generate the TPL feature database.}

\noindent \textbf{App Collection.} 
Based on the selected TPLs, we can choose the corresponding apps. Some apps may include the TPLs from the above two groups at the same time. If there are more in-app TPLs from the first group, we will classify this app as the first group, vice versa.
The first group includes 200 apps and the second group includes 138 apps.
Finally, for the first group of apps, we get 136 TPLs with corresponding 2,413 versions. For the second group of apps, we get 164 TPLs with corresponding 453 versions.

\subsubsection{Dataset for Obfuscation Evaluation.}
In order to evaluate the obfuscation-resilient capability of the tools in a controlled experiment, we need to collect apps both with and without code obfuscation.

For this dataset, we utilize the benchmark from ORLIS~\cite{orlis_benchmark}, which is dedicated to the evaluation of the obfuscation-resilient capabilities. This benchmark is built from 162 open-source apps downloaded from F-Droid~\cite{F-Droid} mapping to 164 TPLs. Our dataset includes two parts: the original non-obfuscated apps and the corresponding obfuscated ones.
We intend to use this dataset to evaluate two aspects of obfuscation-resilient capabilities:
\textit{1) the obfuscation-resilient capabilities towards different obfuscators; 2) the obfuscation-resilient capabilities towards different obfuscation techniques.}

To evaluate the capabilities regarding different obfuscators, each of the 162 apps in the benchmark is obfuscated using three obfuscators (namely \textit{Proguard}~\cite{Proguard}, \textit{DashO}~\cite{Dasho}, and \textit{Allatori}~\cite{Allatori}), respectively. As a result, our dataset includes four sets of apps: a set of 162 non-obfuscated apps, and three sets of apps (162 $\times$ 3) obfuscated by three obfuscators, respectively. These 162 $\times$ 4 apps are provided by ORLIS. However, this provided dataset cannot effectively evaluate the obfuscation-resilient capabilities towards different obfuscation techniques.
Therefore, we extend the dataset from ORLIS by
randomly choosing 88 non-obfuscated apps and using DashO to obfuscate the 88 apps with three different obfuscation techniques (namely control flow randomization, package flattening, and dead code removal). As a result, we include another three sets (88 $\times$ 3) of obfuscated apps in our dataset. We have made our dataset publicly available for the community.




\section{Evaluation}
\label{sec:evaluation}

Our experiments were conducted on 3 servers running Ubuntu 16.04 with 18-core Intel(R) Xeon(R) CPU @ 2.30GHz and {192GB} memory.

\vspace{-1ex}
\subsection{C1: Comparison of Module Decoupling }
\label{sec:evaluation:C1}

{To guarantee the fairness of comparison and effectively distinguish the performance of four module decoupling approaches, we use the same feature extraction method adopted by LibScout to extract code features from TPL candidates and referred TPLs after getting the TPL candidates. We employ the Jaccard index~\cite{jaccard-index} to calculate the similarity between the in-app TPLs and referred TPLs. For this experiment, we know the exact TPL version used by each app, so we consider this decoupling method is effective only when the similarity value is 1. Note that, even if we use the package flattening to obfuscate the apps, this obfuscation technique can keep the code feature unchanged via users' configuration. By comparing the correctly matched number of TPL pairs, we can get the performance of each technique.}


\noindent \textbf{Overall results.}
{TABLE~\ref{tbl:MD-cmp} shows the comparison results of the four module decoupling approaches. According to the result, we can find that the performance from high to low is class dependency (CD), package dependency graph (PDG), homogeny graph (HG), and package hierarchy structure (PHS), respectively.
The PHS as the module decoupling feature is error-prone; it cannot handle all cases in Fig.~\ref{fig:structure}. In contrast, the method of selecting CD as the features outperforms other approaches.}

\begin{table}[t]
\centering 
\caption{The performance comparison of four module decoupling methods.}
\begin{tabular}{l|c|c|c|c|}
\cline{2-5}
\multicolumn{1}{c|}{}                           & \textbf{PHS} & \textbf{HG} & \textbf{PDG} & \textbf{CD} \\ \hline
\multicolumn{1}{|l|}{\textbf{TPL dependencies}}  & \xmark & \moon[scale=0.8]{7.2}    & \moon[scale=0.8]{7.2}   & \moon[scale=0.8]{7.2}    \\ \hline
\multicolumn{1}{|l|}{\textbf{same root package}} & \xmark & \xmark                & \cmark & \cmark \\ \hline
\multicolumn{1}{|l|}{\textbf{obfuscation}}      & \xmark       & \xmark      & \xmark       & \cmark      \\ \hline
\end{tabular}
		\begin{center}
		{\checkmark} \color{black}: can; \xmark\ : cannot; \ \moon[scale=0.8]{7.2} : partially effective;   
		PHS: package hierarchy structure, HG: homogeny graph, PDG: package dependency graph, CD: class dependency   
			
		\end{center}
\label{tbl:MD-cmp}
\vspace{-4ex}
\end{table}


\subsubsection{Reason Analysis}

{If a TPL includes the other dependency TPLs, several independent package trees just form a complete TPL. Nevertheless, PHS cannot handle this case because it considers an independent tree as a TPL instance. Homogeny graph, package dependency graph, and class dependency all consider the class dependency relationship so the three methods can correctly construct the TPL candidates with dependencies. However, Java has different package techniques. If the published TPL is adopted the ``skinny'' method to pack it, the published version will not include the dependency code. For this case, if researchers adopt the original TPLs to build the reference database,
they should consider the package techniques and conduct the module decoupling again to guarantee it can match the exact TPL. However, the existing tools fail to consider the package techniques, which may affect the accuracy. Hence, homogeny graph, package dependency graph, and class dependency graph are partially effective to TPL dependencies.
We suggest that future researchers should pay attention to package techniques.
}

{Different TPLs share the same root package and sub-packages; using PHS as the feature will treat it as one TPLs, leading to inaccurate TPL instances. Homogeny graph also contains the package inclusion relations, which can affect the accuracy of TPL instance construction. Different from the homogeny graph, the package dependency graph is a weighted directed graph. It uses the hierarchy agglomerative clustering algorithm to cut the homogeny graph into different parts based on the intimacy weight. Thus, PDGs can handle different TPLs sharing the same root packages.}


{As for the TPLs with package flattening, the package structure can be changed or completely removed, and PHS, homogeny graph, and package dependency graph include the package inclusion relationship.
Therefore, all of them can be affected by package flattening. In contrast, class dependency does not consider the package, so selecting this feature is totally resilient to package flattening and cannot be affected by package mutation or TPLs sharing the same root package.}

\noindent\fbox{
	\parbox{0.95\linewidth}{
		\textbf{Conclusion to C1:} 
{ Using class dependency in module decoupling stage can achieve better performance.
Most existing tools more or less depend on package name/structure 
as auxiliary features to generate TPL signatures. However, using package structures is extremely unreliable; it is difficult for many tools to find the correct boundaries of in-app TPLs.}
}
}

\subsection{C2: Effectiveness}
\label{sec:evaluation:C2}

We aim to compare existing library detection tools regarding the dataset collected in Section~\ref{sec:datacollection}. Note that LibRadar (a clustering-based method) has published its TPL signature database; we directly employ this database to evaluate its effectiveness. For the remaining tools without the database, we use the collected TPL dataset to build the TPL signature database for each of them. {As we mentioned before, clustering-based tools and some tools only can distinguish whether a package belongs to TPLs or host apps. To ensure the fairness of comparison, we only compare the result at the library-level instead of the version-level for effectiveness comparison.
As for the version-level comparison, we conduct another new experiment and present that in Section~\ref{sec:evaluation:C3}.}


\begin{figure}[t]
	\centering
\includegraphics[scale=0.5]{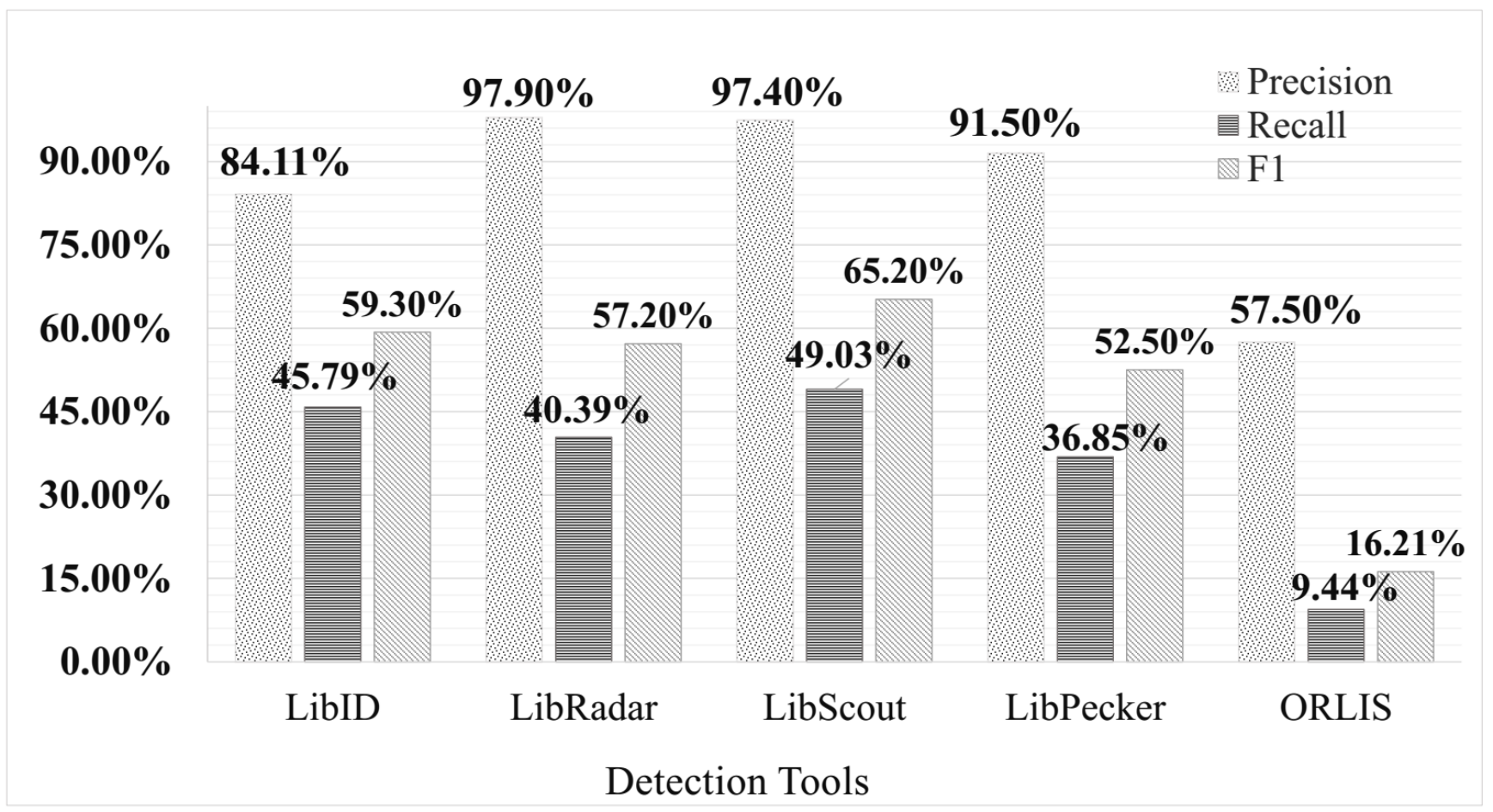}\\
	\vspace{-2mm}
	\caption{Detection result of different TPL detection tools}
	\label{fig:detection_acc}
	\vspace{-2ex}
\end{figure}

\noindent \textbf{Overall Results.}
As shown in Figure~\ref{fig:detection_acc}, it is the detection results of the five publicly available tools.
We can observe that most existing tools can achieve high precision but all tools have low recall (i.e., less than 50\%), indicating that existing tools can only detect less than half of the TPLs used by the apps. LibScout (49.03\%) achieves the highest recall, followed by LibID~(45.79\%).
As for the precision, LibRadar achieves the best performance, which reaches 97.9\%, the precision of LibRadar (97.90\%) and LibScout (97.40\%) and LibPecker (91.50\%) are very close; all of them are above 90\%.
The precision and recall of ORLIS are the lowest among these tools, which are 57.50\% and 9.44\%, respectively. 
To evaluate the comprehensive performance of these tools, we use the F1 value as an indicator. We can see that LibScout outperforms other tools, achieving 65.20\%, followed by LibID (59.30\%). ORLIS has the lowest performance, reaching only 16.21\%.

\subsubsection{Reason Analysis}

We give the reasons that lead to false positives (\textbf{FP}) and false negatives (\textbf{FN}) as follows.

\noindent \textbf{FP Analysis.}
We find that the false positives of existing tools mainly because of 
1) package mutation of the same TPL in different versions, 2) TPL dependency.


{
Even though we just compare the detection results at the library-level, the problem of similar version files is also the primary reason that leads to false positives. Thus, we also discuss the reason in this section.
Our analysis finds that the code of some adjacent versions from the same TPL has high similarity. We calculated their code hash and found some of them even have the same Java code but differences in some native code, resource files, config files, etc. Therefore, it is difficult for existing tools to identify these similar versions, leading to false positives.}

{Some false positives come from package mutation of the same TPL in different versions.
Different versions of the same TPL may have different package hierarchy structures. Even the same TPL, different versions may use different third-party dependencies and some versions may not include the dependencies.
Taking the library ``OkHttp'' as the example, for the versions before 3.0.0, the root package was ``com/squareup/okhttp'', while it changed to ``com/squareup/okhttp3'' for versions after 3.0.0.
LibRadar and LibD cluster the package-level features as the TPL signatures. They will consider that there are two different TPLs during building the TPL reference database since they have different root package structures. For these tools depend on the package hierarchy as a supplementary feature to construct TPL database, if the in-app TPLs are obfuscated and these in-apps have closed versions with high similarity, they may report false positives when they cannot identify the in-app package and can only compare the code similarity. This example also illustrates that these tools cannot identify the root package mutation of the same TPL.
Besides, the code similarities of different versions are also different. The code similarity of some versions is apparent and some are tiny. The diverse code similarity of different versions primarily affects the clustering-based methods. In fact, diversity exists in both versions and different TPLs, which make threshold/parameters selection extremely difficult. It is difficult to select a threshold of the clustering-based method to perfectly distinguish different versions. After getting the clustering result, it also requires manual verification and such a manual process is subject to errors, which may lead to the incorrect reference database.
}



{The other reason is because of TPL dependency.
As can be seen the case 3 from Fig.~\ref{fig:structure}, LIB3 is a TPL that depends on the other TPL LIB1. Considering this situation, there is a LIB3, its version No. is version 2.0, there is the other version LIB3 version 1.0, LIB3 v1.0 does not depend on the LIB1 but the LIB3 v2.0 depends on the LIB1, the code of LIB3 v1.0 and LIB3 v2.0 has high similarity. Given an app, it uses LIB1 and LIB3 v2.0; meanwhile, the TPL feature database also includes all of these TPLs; if a tool uses the package structure to split TPL candidates, it may reports the LIB1 twice, leading to false positives. We can find that it is difficult to find the exact TPL version of LibRadar and LibD because they build the TPL candidate features more or less based on the package structure. 
If a similarity comparison-based method also adopts the package structure to generate potential TPL candidates, it is not reliable as well. From these examples, we also can find that using package structures or package inclusion relations as module decoupling features is error-prone.}



\noindent\textbf{FN Analysis.}
{We take an in-depth analysis of the reasons for the low detection rate of these tools. Compared with the reasons that lead to false positives, the reasons that affect the false negatives are more complicated. We list seven primary reasons that lead to the false negatives here: 1) reverse-engineering tool,2) TPL imported modes, 3) TPL dependency, 4) code obfuscation, 5) TPL identification methods, 6) TPL formats, 7) multidex of Android apps.}

\noindent $\bullet$ \textbf{Reverse-engineering Tool.}
{We find that different reverse-engineering tools have different performances. LibID uses the dex2jar as the reverse-engineering tool; it can transform the jar file into dex format. However, Java and Android are not completely compatible. Java only supported the Android to the version eight; if a TPL is developed by Java 9+, dex2jar cannot identify some new features in this TPL. As a result, the transformation may fail sometimes and directly lead to the TPL features not being generated successfully.
LibID would report some errors when it profiles the TPL files by using dex2jar~\cite{dex2jar}, above reasons lead to 1,106 TPL signatures missing in the database of LibID and lead to the false negatives.
From the success rate of decompilation, we find that Soot outperforms others.}


\noindent $\bullet$ \textbf{ TPL imported mode.} As we mentioned in Section~\ref{sec:import mode}, partial import and customized import can lead to the code of the in-app TPLs is different from the original TPL, which can result in the extracted feature may not match the original ones, leading to false negatives.
For example, we find an app with sha256 prefix ``42A5BC''~\cite{our_lib} includes a popular ad library \texttt{Admob}. This library is partially imported into this app, and it shares the same package name with the host app, i.e., ``de.emomedia''. Some tools cannot find this ad library if they filter the primary module based on the package name (e.g., LibSift), which may exclude some TPLs sharing the same package name with the host app. 
Our analysis finds that all clustering-based tools are impossible to identify the partial import TPLs. Because
the partial imported TPLs of each app may be different, it is difficult to clustering these partial import TPL together. ORLIS only reports the matched packages or classes that belong to TPLs. Therefore, it could identify some partial TPLs. LibPecker also can identify some partial import TPLs because it adopts an adaptive comparison strategy. {Based on our analysis, existing tools have not proposed an effective way to identify partial import TPLs, and we still have a long way to go. Maybe future research can explore this direction and propose a more inspiring method.}

\begin{figure}[t]
	\centering
	\includegraphics[scale=0.4]{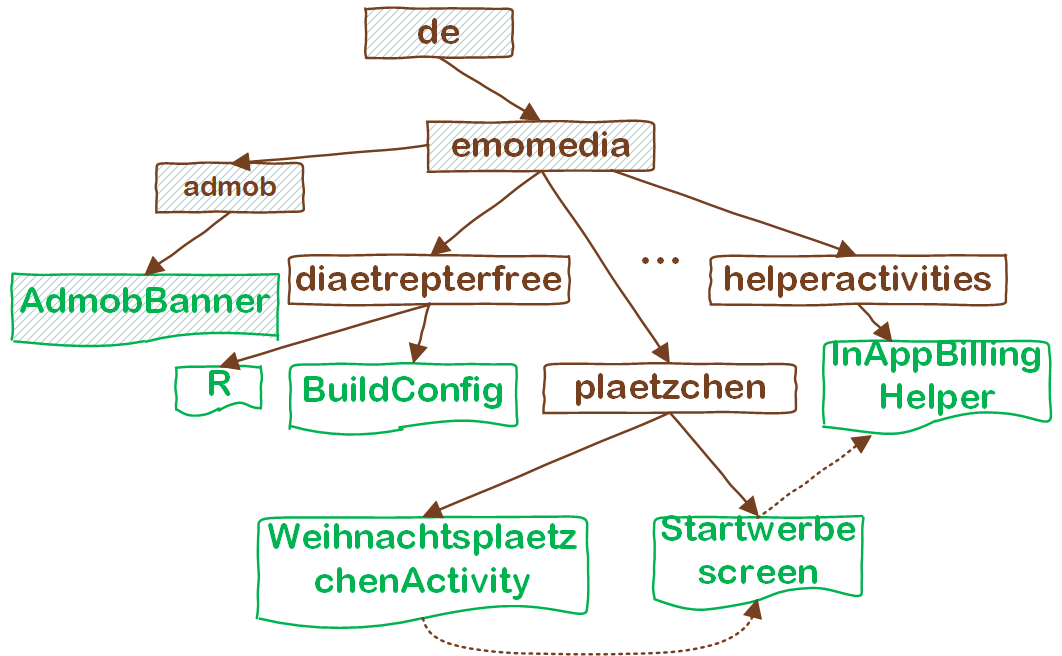}\\
	\caption{An example of a partial import TPL }
	\label{fig:including_example}
	\vspace{-2.5ex}
\end{figure}

\noindent $\bullet$ \textbf{ TPL dependency.}
{Take the case3 in Fig.~\ref{fig:structure} that we present in Section~\ref{sec:challenge:TPL_dependency} as the example; we can find a complete TPL may include some parallel package trees at the root directory. Thus, tools adopt the package structure to split the TPL candidate, which may cut a complete TPL into several different parts. The separated parts may not match the signature of the original TPLs, leading to false negatives.}

\noindent $\bullet$ \textbf{ Code obfuscation.}
Apps in our dataset are from Google Play Store, which may be obfuscated by developers. For example, the dead code removal (a.k.a. TPL Shrink) can delete the code that is not invoked by the host app. Thus, dead code removal can affect all tools. The control flow randomization can change the control flow graph (CFG), which can affect these tools (e.g., LibD, OSSPoLICE) depending on CFG to identify TPLs.
Suppose an obfuscator removes the whole package structure of a TPL and puts all files in the root directory of apps. In that case, this code obfuscation can dramatically decrease the detection rate of tools that include the package structure features. 
Moreover, the package structure can affect the fingerprint generated by LibScout and LibRadar, and the package name mutation can affect the TPL identification of LibID, LibRadar. 
In sum, obfuscation causes FNs of current TPL detection tools.

\noindent $\bullet$ \textbf{ TPL identification methods}
{In addition, the selected identification algorithms may also affect the detection rate.
The detection rate of clustering-based methods primarily relies on the number of collected apps and the reuse rate of in-app TPLs.
It may cause FNs when an insufficient number of apps are collected for clustering. 
Besides, clustering-based tools assume {the modules that are used by a large number of apps are TPLs}~\cite{Wukong2015issta,LibRadar2016ICSE, LibD2017ICSE}. This assumption causes tools using this method can only find some widely-used TPLs.
If some TPLs are seldom used by apps or the number of these apps is very new, they will fail to identify them. Another minor reason for the false negatives of LibRadar is that its pre-defined database may not contain the TPLs we use in our experiment.}
In contrast, the similarity comparison algorithm can alleviate the FNs caused by rarely-used TPLs by adding them to the database.
From our experimental result, we can find the recall of the similarity comparison methods (i.e., LibID, LibScout) is higher than that of clustering-based methods (i.e., LibRadar, LibD).

\noindent $\bullet$ \textbf{TPL formats.}
%
Besides, we find LibID, LibPecker, and ORLIS cannot handle TPLs in ``.aar'' files.
In fact, a TPL could include both the ``.aar'' format files and ``.jar'' format files. In our dataset, about 52\% TPLs are represented in ``.aar'' format. Thus, this is another reason that results in the false negatives of these three tools. 

\begin{figure}[t]
	\centering
	\includegraphics[scale=0.5]{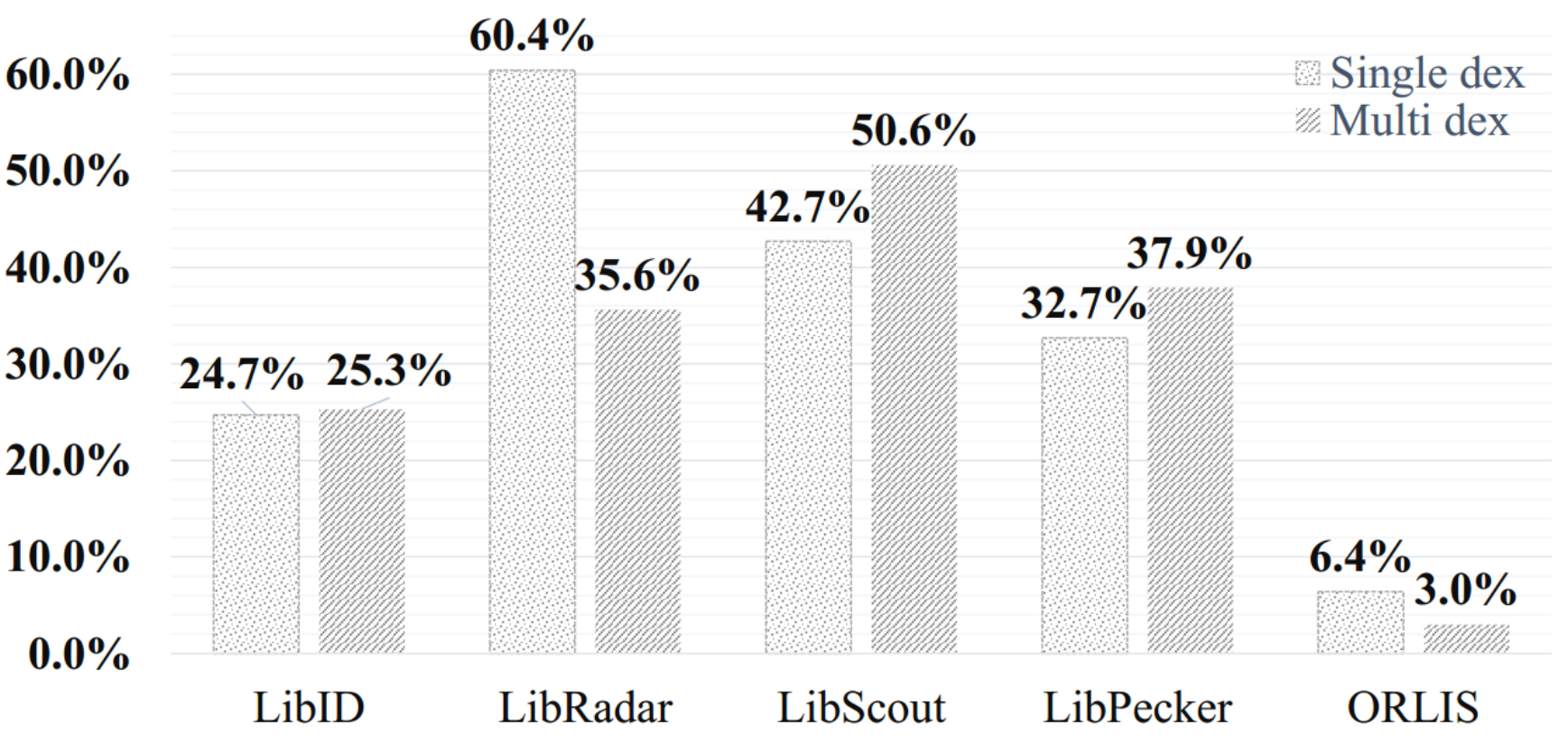}\\
	\vspace{-2mm}
	\caption{Detection rate of different tools towards handling multiple Dex and single Dex problems}
	\label{fig:multidex}
	\vspace{-3ex}
\end{figure}

\vspace{1mm}
\noindent \textbf{Multidex Problem.}
Another reason which can affect the recall is the 64K limit problem.
In Android, the executable Java code is compiled into Dalvik Executable (Dex) file and is stored in the APK file. {The Dalvik Executable specification limits the total number of method references to 64K in a single Dex file.} Nowadays, more and more apps include a large number of TPLs; consequently, it is impossible to include all of these TPL in a single dex file.
Whereas, since Android 5.0 (API level 21), it supports compiling an app into multiple Dex files if it exceeds the limited size, which allows the extra TPLs could scatter in these dex files

We try to figure out whether these state-of-the-art TPL detection tools have considered these new features or they just can handle the single dex file. Accordingly,
we refer to the source code of these publicly-available tools and find that LibRadar and ORLIS only consider the analysis on a single Dex file, which may cause the missing of a considerable amount of TPLs in detection.
The specific detection result of these tools to handle the single dex and Multidex is shown in Figure~\ref{fig:multidex}. The result also indicates that when LibRadar and ORLIS deal with the single dex, the detection rate of LibRadar and ORLIS are 60.4\% and 38.6\%, respectively. However, when they handle the apps with multi-dex, the detection rate decreases by almost half than the original one.

We further investigate the percentage of such multiple Dex apps in our evaluation dataset. We find that only 41 apps contain a single Dex file, and the remaining 180 apps contain multiple Dex files, and more than 80\% of them contain three Dex files. Surprisingly, the app ``com.playgendary.tom'' even contains 98 Dex files. For example, ``bubbleshooter.orig'' uses 19 TPLs, but the classes.dex file only contains 4 TPLs of them; the remaining libraries are included in other \texttt{classes\_N.dex} files. Without a doubt, if the tool just considers the single Dex situation, its effectiveness (especially the recall) will be significantly decreased.
Therefore, we suggest that TPL detection tools should take into account multiple Dex situations to ensure reliable results.

{Based on our analysis, we find that apart from the challenges in TPL detection itself, many issues that affect the detection performance were mainly ignored by previous researchers. For example, LibPecker sets a strong assumption that assumes the package structure can be preserved during obfuscation. LibID assumes that most obfuscators cannot affect the inter package hierarchy structure. These strong assumptions directly limit their algorithm design and finally affect their detection performance.  Fortunately, these problems ignored by previous researchers can be fixed easily. We will elaborate on our implications about how to improve the performance of TPL detection tools in Section~\ref{sec:discussion}.}

\vspace{2mm}
\noindent\fbox{
	\parbox{0.95\linewidth}{
		\textbf{Conclusion to C2:} 
{As for effectiveness, LibScout performs the best while ORLIS gets the lowest performance among these tools. Most tools achieve high precision but low recall because these tools ignore the challenges that we mentioned in Section~\ref{Sec:challenge} such as package flattening techniques, TPL dependency and complicated package structure, leading to a substantial number of false negatives. Besides, some previous methods may ignore some updating features of apps and TPLs, such as the new format of TPLs and the new compilation method. }
}}

\subsection{C3: Efficiency}
\label{sec:evaluation:C3}

{Considering efficiency, we use the detection time as the metric to evaluate each tool.
To ensure a fair comparison, we first employ each tool to generate TPL features for similarity comparison-based tools. The detection time does not include the TPL feature database generation time, which is the practice in all related work. The time cost consists of the TPL detection process for an app, i.e., pre-process an app, profile the app, module decoupling, extract the code feature, identify TPLs inside. For each tool, the detection time mainly depends on the number of features in a database and the number of TPLs in each app.}

\noindent{\textbf{Results.}} 
Table~\ref{tbl:detection_time} shows the detection time of the five selected tools. We can find that the efficiency of different tools has big differences.
Several tools may just cost a few seconds but others may cost several hours.
The average detection time of LibRadar is 5.53s, which is the fastest one among these tools, followed by LibScout (82.24s). The detection time of LibID and LibPecker is much longer; the median detection time of these two tools is more than \textbf{4} hours for each app. 
Surprisingly, the average detection time of LibID is even nearly one day per app. We find that if the size of a dex file of a TPL is larger than 5MB, the detection time of LibID dramatically increases. 
Based on our observation, we find that LibID cannot handle too many TPLs simultaneously and is computation heavy in terms of CPU and memory. For each detection process, LibID needs to load all the features of TPLs in the memory, and it costs about 21 minutes to load all data.
Furthermore, the average detection time of ORLIS is 1438.50s per app.
These three tools are time-consuming, which may not be suitable for large-scale TPL detection.

\begin{table}[t]
	\centering
	\small
	\caption{Detection time of different tools}
    \vspace{-3mm}
 \scalebox{0.9}{
	\begin{tabular}{cccccc}
		\toprule[1.3pt]
		\textbf{Tool} & \textbf{LibID} & \textbf{LibRadar} &  \textbf{LibScout} & \textbf{LibPecker} & \textbf{ORLIS}     \\
		\midrule[0.8pt]
		Q1  &  2.07h &   5s      & 47s    &   {3.38h}  &  876.75s      \\
		\midrule
		Mean  &  23.12h & 5.53s   &  82.42s  &  {5.11h}    & 1438.50s  \\
		\midrule
		Median  & 6.56h &  5s   &    66s      & {4.65h}  & 1199.5s    \\
		\midrule 
		Q3     & 20.04h & 6s &    95.25s   &   {6.46h}    &   1571.5s  \\
		\bottomrule[1.3pt]
	\end{tabular}
}	
	\label{tbl:detection_time}
	\vspace{-2ex}
\end{table}

\subsubsection{Reason Analysis}


{For the above experimental results, we will analyze the reasons according to each stage of the TPL detection process, and we summarized {four} factors affecting the efficiency of different tools as follows.}

\noindent \textbf{Reverse-engineering tools.} 
In the preprocessing stage, apart from LibRadar, the efficiency of the reverse-engineering tools used by other tools is similar.
LibRadar can directly process the \texttt{.dex} files by using their tool \textsc{libdex}~\cite{LibRadar_tool} that dramatically improves the performance. 
Besides, as we mentioned in Section~\ref{sec:evaluation:C1}, {LibRadar only handles the single Dex file and ignores the multiple Dex problem, however, more than two-thirds of apps contain multiple dex files in our dataset, which is also one of the reasons that LibRadar is faster than other tools. 
However, the preprocessing is not the main reason that affects system efficiency.} 

\noindent \textbf{Extracted Features.}
The complexity of extracted features also can affect time consumption. In general, it takes more time to extract the semantic information than to extract the relatively simple syntax information.

{LibRadar extracts the framework Android APIs and LibScout extracts the fuzzy method signatures as the code signatures. Compared with other tools, these code features are relatively simple because they just consider the syntactic information. Thus, their detection efficiency is better than other tools.
Both ORLIS and LibScout choose the fuzzy method signature as the code features to represent TPLs, but ORLIS also extracts the method invocation relationship and class inheritance relationship as the code features, which adds the extra computational complexity, leading to longer time consumption than LibScout. LibID, LibPecker, and ORLIS all consider the class dependency, and extract these features usually time-consuming.} 
Among these tools, the most time-consuming tool could be OSSPoLICE.
OSSPoLICE uses the CFG centroid as the code feature. However, to get the CFG centroid, OSSPoLICE needs to traverse the CFG, traversing a graph is a time-consuming process. The specific analysis we will give in comparison strategy part.

 \begin{figure*}[t]
	\begin{minipage}{0.33\textwidth}
		\centering
		\includegraphics[scale=0.39]{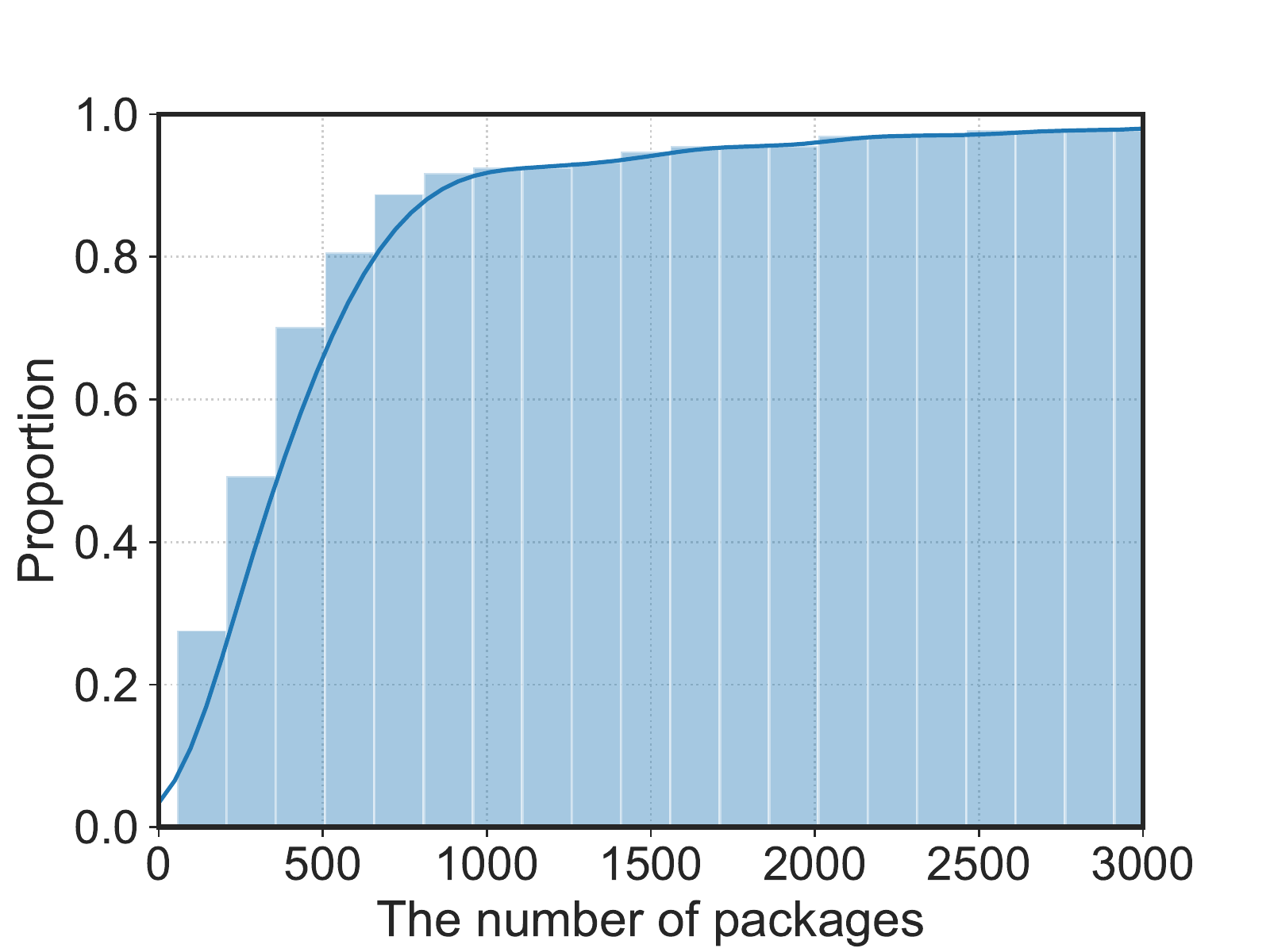}
		\vspace{-4mm}
		\label{fig:package_cnt}
	\end{minipage}%
	\begin{minipage}{0.33\textwidth}
		\centering
		\includegraphics[scale=0.39]{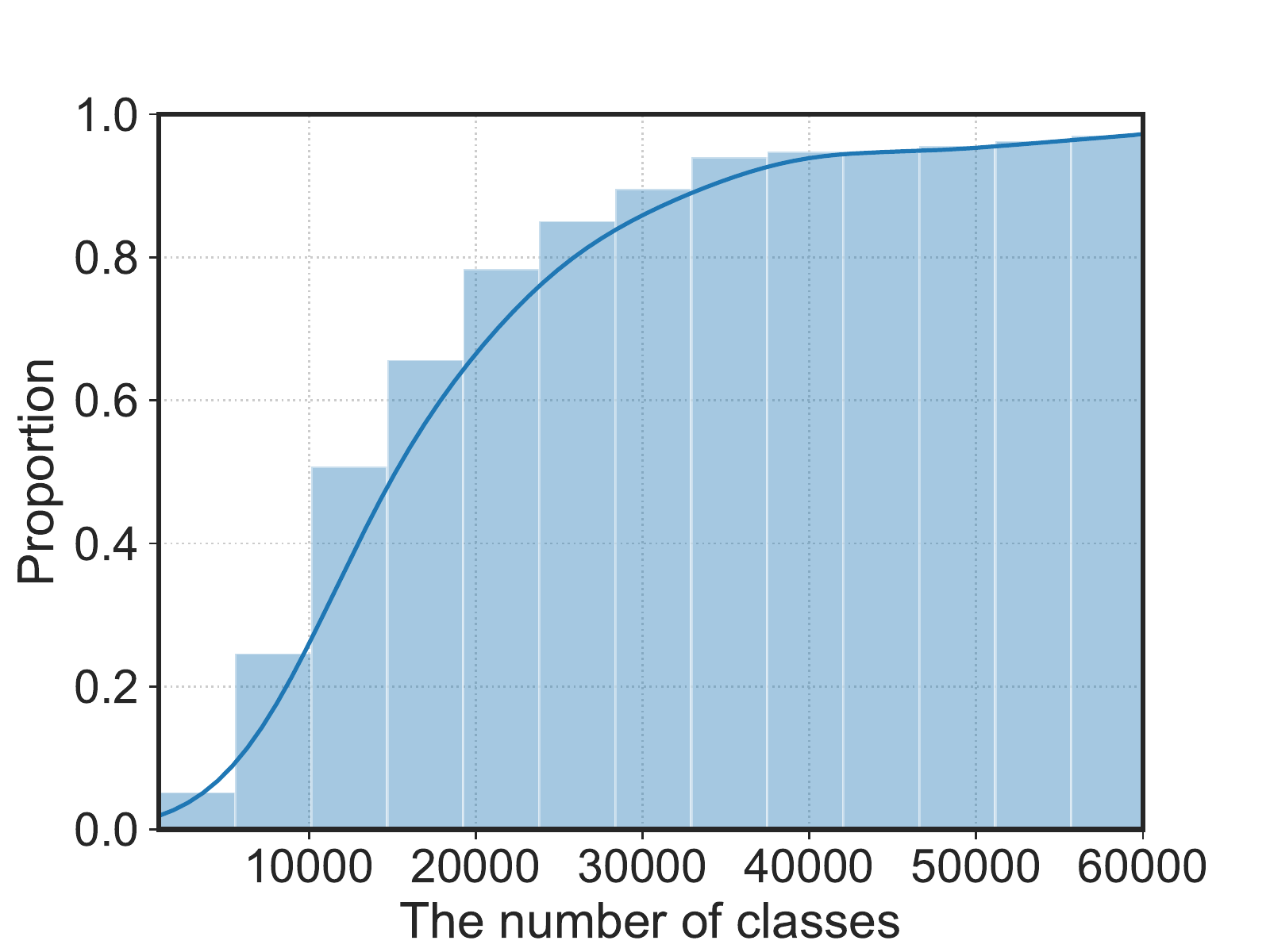}
		\vspace{-4mm}
	\end{minipage}%
	\vspace{-1ex}
		\begin{minipage}{0.33\textwidth}
		\centering
		\includegraphics[scale=0.4]{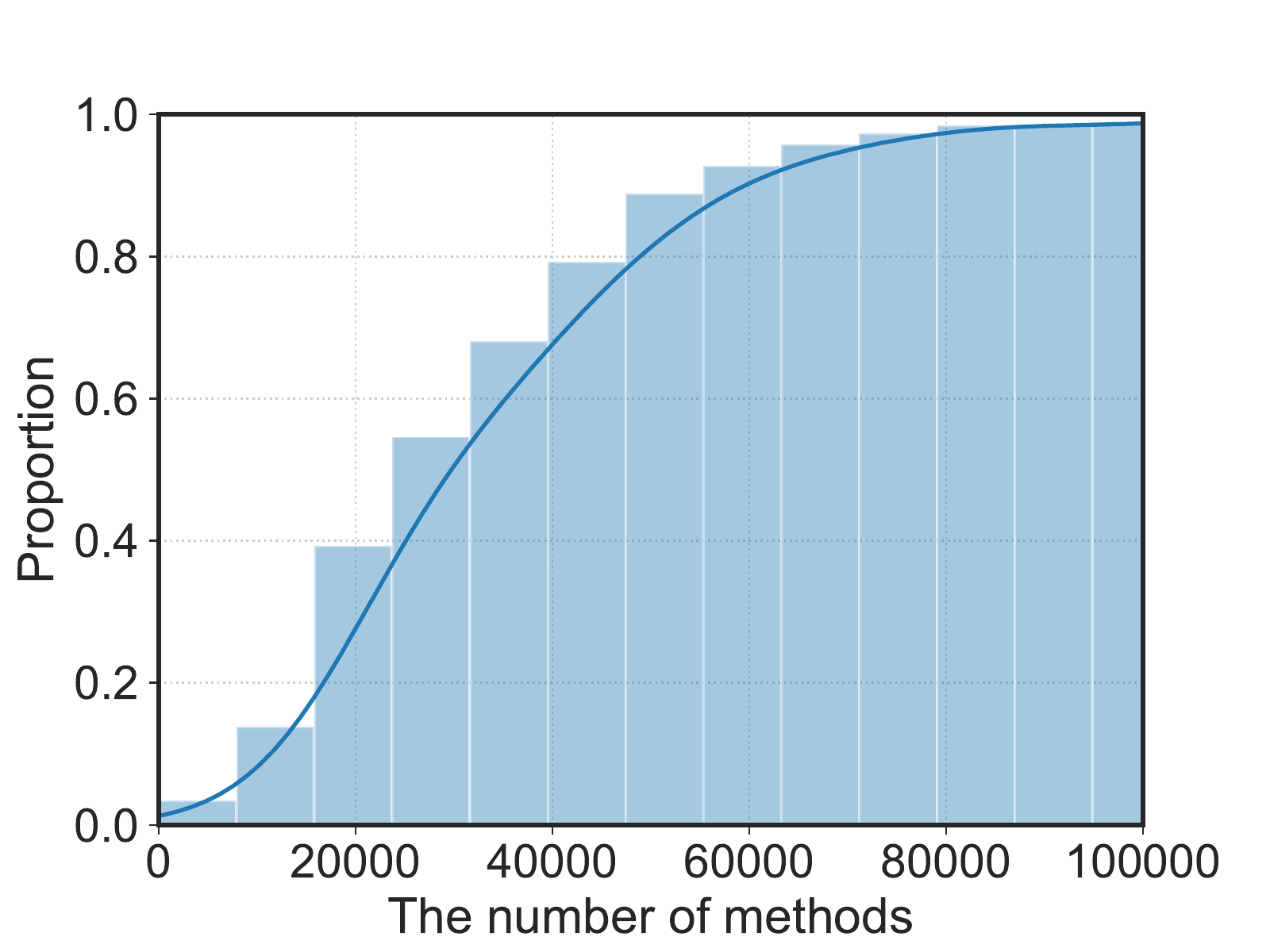}
		\vspace{-4mm}
	\end{minipage}%
	\caption{CDFS of the number of packages, classes and methods in our dataset}
	\label{fig:CDF}
	\vspace{-2ex}
\end{figure*}

\noindent \textbf{Feature Granularity.}
We also find that the granularity of TPL code feature can affect detection efficiency. Currently, there are two levels of granularity used by existing TPL detection tools: package-level (i.e., LibRadar and LibScout) and class-level (i.e., LibID, ORLIS, and LibPecker). 

To explain how feature granularity affects the detection efficiency, we counted the number of packages, classes, and methods of all apps in the ground truth. Fig~\ref{fig:CDF} illustrates the Cumulative Distribution Functions (CDFs) of the number of packages, classes and methods; we can find that  the number of package-level items is far less than that of the class-level. The number of packages of 95\% of apps is less than 1000, and the number of classes of 95\% of apps is large than 10K. The average number of classes is about 40 times as much as the number of packages.
Based on our experimental result, we can also observe that the overhead of systems that use class-level features is higher than systems using package-level features. 
Thus, the efficiency of LibID, LibPecker, and ORLIS is obviously worse than other tools. If the functionality of a TPL is complicated, the number of classes and methods could also increase, thus the comparison time increases accordingly.
The growth rate of comparison time between the lib-class and app-class is exponential. Thus, we can find that the average detection time of LibID, LibPecker and ORLIS is much longer than the median time. This is because it takes more time to extract code features of some complicated TPLs.
Take LibPecker as an example; even if an app contains just one TPL, one comparison could cost time ranging from about 5s to 10s, and there are 2,115 TPL features to be compared in our dataset. The average detection time could reach 5.11 hours.

\noindent \textbf{Comparison Strategy.} 
{For most tools, the feature comparison strategy is the step that has the greatest impact on the efficiency of TPL detection tools.}
%

\begin{table}[t]
  \centering
  \caption{The time complexity of the state-of-the-art TPL detection tools }
  \vspace{-2ex}
  \scalebox{0.95}{
  \begin{tabular}{cc}
    \toprule[0.4mm]
    \textbf{Tool} & \textbf{Time Complexity} \\
    \midrule
    \midrule
     LibRadar & $O(p \times n \times \overline{p})$ \\
        \rowcolor[rgb]{ .937,  .965,  .918} LibScout & $ O(p \times n \times \overline{p}) + O(c \times \overline{c}) $ \\
    \rowcolor[rgb]{ .886,  .937,  .855} OSSPoLICE & $  O(cm(n'+e)(c'+1)) +  O(p\times n\times\overline{p}) + O(c\times\overline{c}) $  \\
    \rowcolor[rgb]{ .776,  .878,  .706} ORLIS &$O(c \times n \times \overline{c})$ \\
    \rowcolor[rgb]{ .663,  .816,  .557} LibPecker &$O(p \times n \times \overline{p})+ O(c \times \overline{c}) + O(c!)$ \\
    \rowcolor[rgb]{ .439,  .678,  .278} \textcolor[rgb]{ 1,  1,  1}{LibID} & \textcolor[rgb]{ 1,  1,  1}{$O(p \times log(c) \times log(\overline{c})\times n \times \overline{p})+ O(p \times \overline{p}) $} \\
    \bottomrule[0.4mm]
  \end{tabular}
  }
  \label{tbl:time_cmplex}
  \vspace{-3ex}
\end{table}

{To explain the reasons more clearly, we present the time complexity of each tool in Table~\ref{tbl:time_cmplex}. Given a test app, the number of the packages (\textbf{P}) is denoted as $p$; the number of classes in this apps is denoted as $c$ and number of methods in this apps is denoted as $m$. According to the Fig~\ref{fig:CDF}, we can get $m>c>p$ in general situations. Suppose there are $n$ TPLs stored in the feature database. We use $\overline{p}$ to represent the number of the packages of potentially matched libraries in the database. We use $\overline{c}$ to denote the number of classes of a library in the database. The number of methods of a library in the database is denoted as $\overline{m}$.}

{For LibRadar, it compares the library feature at the package-level. Thus, the time complexity is $O(p \times n \times \overline{p})$. For LibScout, it first compare the package-level similarity but if it cannot find the exact matching, it will compare the class-level similarity, hence, the complexity is $O(p \times n \times \overline{p}) + O(c \times \overline{c})$.
OSSPoLICE is a TPL version identification tool, it first identifies the used TPL range and then ensures the specific version files.
Therefore, OSSPoLICE and LibScout have the same time complexity in comparison stage, but OSSPoLICE could cost substantial time in feature extraction stage. 
OSSPoLICE use centroid to denote a CFG. The CFG centroid is a three-dimensional vector. The first dimension represents the node index, the second dimension represents the outgoing degree, the third dimension indicate the circle in the graph. Getting the third dimension data could be very time-consuming.
Traversing graph is a time-consuming process. The time complexity of this process is $O(cm(n'+e)(c'+1))$. $O(n'+e)$ is the complexity to find all the loops in a graph, where there are n' nodes, e edges and c' elementary in a graph~\cite{circle}. 
ORLIS does not consider the package structure, therefore the time complexity is $O(c \times n \times \overline{c})$. LibPecker adopts the similar comparison strategy with LibScout, but it adds the adaptive class matching to improve the accuracy. LibPecker also calculates the weight for each class, thus, it needs extra time consumption $O(c!)$.
For LibID, it formulates the class pairs match as the Binary Integer Programming~\cite{BIP} problem, the time complexity is $O(p \times log(c) \times log(\overline{c})\times n \times \overline{p})$. When LibID gets the matched class pairs, it also need to calculate the matched package pairs, the time complexity is $O(p \times \overline{p})$.}

Based on the equations of the time complexity, we can sum up the comparison order into two types: the top-down comparison strategy (from package-level features to class-level features), such as LibScout and LibPecker; the other type is bottom-up comparison strategy (from class-level features to package-level feature), such as LibID and ORLIS. For the comparison strategy, we recommend top-down strategy, consider the time complexity $O (p \times n \times \overline{p}) + O(c \times \overline{c})$ is obviously smaller than $  O(c \times \overline{n} \times \overline{c}) +  O( {p} \times \overline{p})$.


\vspace{2mm}

\noindent\fbox{
	\parbox{0.95\linewidth}{
		\textbf{Conclusion to C3:} 
{LibRadar outperforms other TPL detection tools regarding the detection time, taking only {5.53s}/app on average. LibID takes more time than others, almost one day per app.
The efficiency of a TPL detection tool is primarily determined by the extracted features, the feature granularity and the comparison strategy.}
}}

\subsection{C4: Capability of version identification.}	
\label{sec:evaluation:C4}
{ For version identification, we set a stricter rule compared with C2. For example, given a test app including a TPL \textit{Okio 1.2.0}, if the detection result of a tool is \textit{Okio 1.2.0} and \textit{Okio 1.3.0}, we consider there is a false positive (i.e., \textit{Okio 1.3.0}) and a true positive (i.e., \textit{Okio 1.2.0}). If the detection result is \textit{Okio 1.3.0} and \textit{Okio 1.4.0}, we consider there are two false positives (both \textit{Okio 1.3.0} and \textit{Okio 1.4.0} are incorrect) and one false negatives (the tool does not identify the correct version, \textit{Okio 1.2.0}). For library-level identification, we only consider the tool finds the used TPL without specifying the version. For version-level identification, tools need to identify the exact version of used TPLs.
{We use false positives (FP) and false negatives (FN) precision, recall and F1-value to evaluate existing tools.}
Note that, if two versions of the same TPL have the same Java code and a tool reports the two versions at the same time, we do not consider that is a false positive.
}  
{Note that not all tools compared in C1 have the capability of version identification, such as the clustering-based method. Thus, we remove LibRadar and ORLIS.
As we analyzed in previous sections, LibRadar can only construct the TPL features at the package-level. Nevertheless, it is challenging to distinguish the differences among different versions by using package-level features. For one thing, package mutation of different versions is very common~\cite{libscout2016ccs}. For another thing, the code differences of different versions are also various. The code features among some versions may not be so obvious. 
The {parameter configuration} is also a big challenge for clustering algorithms. Meanwhile, it also requires substantial manual verification to ensure the correct versions for clustering-based methods. However, the authors of LibRadar only manually verified about 200 TPLs at present. Therefore, some TPLs in our benchmark may not be verified by LibRadar, and this is another reason we do not compare it.
ORLIS is a similarity comparison-based tool, but it presents the final result just by reporting the classes that belong to TPL and classes of different versions could be the same, so we cannot verify the specific versions. Hence, we do not compare the two tools here.
}

\begin{table}[t]
  \centering
  \caption{Comparison results of existing TPL detection tools regarding the version identification}
  \vspace{-1ex}
\scalebox{0.83}{
\begin{tabular}{c|ccc||ccc|}
\cline{2-7}
 & \multicolumn{3}{c||}{\textbf{Group 1}} & \multicolumn{3}{c|}{\textbf{Group 2}}\\ \cline{2-7} 
 & \textbf{LibID} & \textbf{LibScout} & \textbf{LibPecker} & \textbf{LibID} & \textbf{LibScout} & \textbf{LibPecker} \\ 
 \hline
 \Xhline{1pt}
\multicolumn{1}{|c|}{FP} & \cellcolor[HTML]{ECF4FF}\textit{\textbf{123}} & \cellcolor[HTML]{ECF4FF}134 & \cellcolor[HTML]{ECF4FF}507 & \cellcolor[HTML]{ECF4FF}\textit{\textbf{344}} & \cellcolor[HTML]{ECF4FF}6065 & \cellcolor[HTML]{ECF4FF}9235 \\
\multicolumn{1}{|c|}{FN} & 533 & 291 & 268 & 685 & 88 & 10 \\
\multicolumn{1}{|c|}{Precision} & \cellcolor[HTML]{ECF4FF}\textit{\textbf{69.34\%}} & \cellcolor[HTML]{ECF4FF}64.66\% & \cellcolor[HTML]{ECF4FF}59.96\% & \cellcolor[HTML]{ECF4FF}\textbf{46.43\%} & \cellcolor[HTML]{ECF4FF}12.72\% & \cellcolor[HTML]{ECF4FF}9.46\% \\
\multicolumn{1}{|c|}{Recall} & 34.25\% & 64.29\% & 60.63\% & 11.70\% & 88.73\% & 98.72\% \\
\multicolumn{1}{|c|}{F1-value} & 45.85\% & 64.47\% & 60.29\% & 18.69\% & 22.25\% & 17.27\% \\
\Xhline{1.5pt}
\end{tabular}
}
  \label{tbl:version_identify}%
  \vspace{-4ex}
\end{table}

\noindent \textbf{Overall Result.} 
For this experiment, we estimate the version identification capability of state-of-the-art TPL version detection tools. Besides, we also evaluate the performance of these tools towards different version similarities. To this end, we set two groups of datasets. {Compared with the code similarity of different versions of TPLs in Group 1, Group 2 is more obvious.}
TABLE~\ref{tbl:version_identify} is the comparison results of TPL detection tools regarding version identification.
The version identification capability is mainly reflected by false positives and precision. Compared with the result in Fig.~\ref{fig:detection_acc}, we can find the precision of all tools has decreased from TABLE~\ref{tbl:version_identify}. The number of false positives of all tools is not a minority for both datasets.
%
%
LibID always outperforms others in version identification even though the adjacent versions of TPLs have high code similarity. In contrast, the version identification capability of LibPecker is the worst based on precision. Especially for the second dataset (the code features of versions of the same TPL have high similarity), the precision of LibScout and LibPecker are dramatically decreased; each of them is 12.72\% and 9.46\%, respectively.
{The detection results show that the detection results of all tools include many imprecise identifications, which means we have a long way to go for precise version identification.}


\subsubsection{Reason Analysis}
For version identification capability, we can find that all tools report many false positives.
Based on our analysis, two reasons are leading to this result, i.e., 1) feature granularity and 2) similarity threshold.

The primary reason is that the \textit{feature granularity} of the code features extracted by existing tools are too coarse to distinguish the code changes of different versions.
That is a common issue for all existing tools. 
For example, LibScout uses the fuzzy descriptor (a variant of method signature) as the birthmark to represent TPLs. However, it is impossible for this feature to find the modifications of statements in the same method of different versions, leading to false positives. Besides, even the different method signatures may generate the same fuzzy descriptor, which cannot effectively distinguish the tiny changes between two adjacent versions.
{For instance, there are two different method signatures: int methodA (classA, bool, com.amazon.session) and int methodB(com.chicho.sessionCallback, bool, classB); By using the method of LibScout, the two different method signatures will get the same signature, i.e., int X(X, bool, X).}
Choosing CFG or CFG centroid as the feature also cannot detect the changes of statements in a method. LibRadar uses the system-defined APIs as the code feature, it can be resilient to renaming obfuscation but it cannot find the changes of developer-defined APIs.
%
Both LibID and LibPecker choose the class dependency as the code feature; they also have the same problem.
But LibID also contains extra features, such as the fuzzy method signatures and CFGs. That is to say, apart from the CFG, LibID also contains the features of LibScout and LibPecker. This is why LibID has better precision in identifying similar versions.



{Apart from the feature granularity, the second reason is due to the selection of similarity threshold. The test datasets of previous tools and ours are different. The optimal threshold of the comparison algorithm on different datasets may be different.}

Moreover, we use a hash function sha256 to calculate hash values for the Java code of the TPLs in our dataset and find that some versions of the same TPL actually have the same hash value, which means they have the same Java code but with different native code, resources or config files. Existing tools only consider the Java code of TPLs but ignore other attributes of TPLs. Future researchers may consider using other attributes to distinguish different versions, which may increase the version identification precision. 

{Compared the results of LibScout and LibPecker with the results library-level identification in Fig.~\ref{fig:detection_acc}, we can find the recall increased in this experiment. Two reasons cause this result:
1) we use the commercial apps from Google Play and some of in-app TPLs are obfuscated by developers, while the open-source apps in this experiment are not obfuscated. Obfuscation also can affect the recall.
}

\vspace{1ex}
\noindent\fbox{
	\parbox{0.95\linewidth}{
		\textbf{Conclusion to C4:} 
The state-of-the-art TPL detection tools can identify the in-app TPLs with high precision. However, there is still a long way to go for the exact version identification of current tools. The version detection results of existing tools usually include many FPs.
}}

\subsection{C5: Obfuscation-resilient Capability}
\label{sec:evaluation:C5}
In this section, we attempt to investigate the obfuscation-resilient capability of existing tools from two aspects: (1) towards different obfuscation tools; and (2) towards different obfuscation strategies.
We also use precision, recall and F1-value to evaluate existing tools (i.e., LibID, LibPecker, ORLIS, LibRadar and LibScout).

\subsubsection{\textbf{Evaluation towards Different Obfuscators.}} 
\label{eval_obfuscators}
Users can configure the obfuscation strategies of obfuscators by themselves. The obfuscation strategies of the three obfuscators are shown in Table~\ref{tbl:obfuscation_config}. We can see that
different obfuscation tools support different obfuscation strategies and have different obfuscation capabilities.
\textit{Proguard} only enables two strategies, including identifier renaming and package flattening while \textit{DashO} enables all of the listed strategies.
We compare the detection result of the five tools on apps with/without being obfuscated by different obfuscation tools, in an attempt to investigate their effectiveness.

\begin{table}[t]
	\centering
	\small
	\caption{Enabled obfuscation strategies of each obfuscator}
	\vspace{-3mm}
	\scalebox{0.9}{
	\begin{tabular}{lccc}
		\toprule
		\textbf{Obfuscation strategy}	          & \textbf{Proguard}              & \textbf{Allatori}        & \textbf{DashO}       \\
		\midrule
		\rowcolor{gray!20}
		Dead Code Removal      & \xmark                & \xmark          &  \color{green}\cmark \\
		String Encryption & \xmark                & \color{green}\cmark & \color{green}\cmark  \\
		\rowcolor{gray!20}
		Control Flow Randomization & \xmark       & \color{green}\cmark & \color{green}\cmark  \\
		Identifier Renaming & \color{green}\cmark & \color{green}\cmark & \color{green}\cmark  \\
		\rowcolor{gray!20}
		Package Flattening & \color{green}\cmark  & \color{green}\cmark &\color{green}\cmark  \\
		\bottomrule
	\end{tabular}
	}
	\begin{center}
		\color{green}{\cmark} \color{black}: enabled \xmark: disabled
	\end{center}
	\vspace{-2ex}
	\label{tbl:obfuscation_config}
	\vspace{-2ex}
\end{table}

\noindent $\bullet$ \textbf{Results.}
{We first compare the comprehensive performance by using F1-value as the metric to evaluate each tool. The results are showed in TABLE~\ref{tbl:f1-value_tools_obf}. The shade of blue cell indicates the degree of decrease and the orange cell indicates the degree of increase in TABLE~\ref{tbl:f1-value_tools_obf}.
}
{For apps without obfuscation, LibScout outperforms others, reaching 83.67\%, followed by LibRadar 77.09\%.
We can find that the results are also consistent with the result of C2, even though there are two different datasets.
We can note that the performance of LibID in C2 and C5 has orders of magnitude of differences. The recall of LibID dramatically dropped here. As we mentioned before, the performance of LibID is greatly limited by dex2jar~\cite{dex2jar}.
Apart from the compatibility issues, dex2jar also has a strict type detection mechanism but obfuscation can change the variables' type, leading to LibID not decompiling these obfuscated apps successfully. These reasons can discount the recall of LibID and lead to the inconsistent results of LibID in C2 and C5.}
{For apps with obfuscation, we can find the performance of all tools for apps without obfuscated keeps the same as the apps with obfuscated by Proguard, indicating that all tools can effectively identify the TPLs in apps obfuscated by Proguard. Furthermore, based on the TABLE~\ref{tbl:obfuscation_config}, we can find that Proguard only implements the identifier renaming and package flattening. The package flattening of Proguard just changes the package name obfuscation. We know all tools extract the code features without containing the unstable identifiers based on our previous analysis. The results also proved that all tools are resilient to renaming obfuscation.
However, they all fail to effectively detect TPLs obfuscated by Dasho, leading to a sharp decline in detection rate. LibRadar is the worst, the performance dropped by 63.97\%, followed by LibScout, its reduce to 35.96\% (dropping by 47.71\%)
The code obfuscation effect of Allatori lies between Proguard and DashO. Here,  we can find that the detection rate of LibPecker increased. We will analyze the specific reasons below.}


\begin{table}[t]
	\centering
	\small
\caption{Comprehensive performance (F1-value) comparison of existing tools regarding the resiliency to different obfuscation tools}
	\vspace{-2ex}
\begin{tabular}{ccccc}
		\toprule
			\multirow{2}{*}{\textbf{Tool}} & \multirow{2}{*}{\begin{tabular}[c]{@{}c@{}}\textbf{Without}\\ \textbf{obfuscation}\end{tabular}} & \multicolumn{3}{c}{\textbf{With Obfuscation}} \\ \cline{3-5}
			&     & \textbf{Proguard}     & \textbf{Allatori}  &  \textbf{Dasho}   \\ 
			\midrule
LibID & 18.69\% & \cellcolor{gray!20}18.69\% & 14.20\% & 10.21\% \\
LibPecker & 69.05\% & \cellcolor{gray!20}69.05\% & \cellcolor{orange!40}\textbf{76.21\%} & 34.88\% \\
ORLIS & 62.31\% & \cellcolor{gray!20}62.31\% & 60.99\% & 34.70\% \\
LibRadar & 77.09\% & \cellcolor{gray!20}77.09\% & 75.85\% & \cellcolor[HTML]{DAE8FC}\textbf{13.12\%} \\
LibScout & 83.67\% & \cellcolor{gray!20}83.67\% & \cellcolor[HTML]{ECF4FF}\textbf{42.53\%} & \cellcolor[HTML]{DAE8FC}\textbf{35.96\%} \\ 
\bottomrule[0.9pt]
\end{tabular}
\label{tbl:f1-value_tools_obf}
\vspace{-2ex}
\end{table}

\begin{table*}[t]
	\caption{Comparison results of state-of-the-art TPL detection tools regarding the code obfuscation-resilient capability towards different obfuscators}
	\vspace{-2ex}
\begin{tabular}{ccccccccccccc}
\Xhline{1.3pt}
\multicolumn{1}{c|}{} & \multicolumn{6}{c|}{\textbf{Precision}} & \multicolumn{6}{c}{\textbf{Recall}} \\ \cline{2-13} 
\multicolumn{1}{c|}{} & \multicolumn{2}{c|}{\textbf{Proguard}} & \multicolumn{2}{c|}{\textbf{Allatori}} & \multicolumn{2}{c|}{\textbf{DashO}} & \multicolumn{2}{c|}{\textbf{Proguard}} & \multicolumn{2}{c|}{\textbf{Allatori}} & \multicolumn{2}{c}{\textbf{DashO}} \\ \cline{2-13} 
\multicolumn{1}{c|}{} & \multicolumn{1}{c|}{} & \multicolumn{1}{c|}{} & \multicolumn{1}{c|}{} & \multicolumn{1}{c|}{} & \multicolumn{1}{c|}{} & \multicolumn{1}{c|}{\textbf{changed}} & \multicolumn{1}{c|}{} & \multicolumn{1}{c|}{} & \multicolumn{1}{c|}{} & \multicolumn{1}{c|}{\textbf{changed}} & \multicolumn{1}{c|}{} & \textbf{changed} \\
\multicolumn{1}{c|}{\multirow{-4}{*}{\textbf{Tool}}} & \multicolumn{1}{c|}{\multirow{-2}{*}{\textbf{ratio}}} & \multicolumn{1}{c|}{\multirow{-2}{*}{\textbf{\begin{tabular}[c]{@{}c@{}}changed\\ rate\end{tabular}}}} & \multicolumn{1}{c|}{\multirow{-2}{*}{\textbf{ratio}}} & \multicolumn{1}{c|}{\multirow{-2}{*}{\textbf{\begin{tabular}[c]{@{}c@{}}changed\\ rate\end{tabular}}}} & \multicolumn{1}{c|}{\multirow{-2}{*}{\textbf{ratio}}} & \multicolumn{1}{c|}{\textbf{rate}} & \multicolumn{1}{c|}{\multirow{-2}{*}{\textbf{ratio}}} & \multicolumn{1}{c|}{\multirow{-2}{*}{\textbf{\begin{tabular}[c]{@{}c@{}}changed\\ rate\end{tabular}}}} & \multicolumn{1}{c|}{\multirow{-2}{*}{\textbf{ratio}}} & \multicolumn{1}{c|}{\textbf{rate}} & \multicolumn{1}{c|}{\multirow{-2}{*}{\textbf{ratio}}} & \textbf{rate} \\ \hline
\midrule[0.6pt]

LibID & \cellcolor[HTML]{68CBD0}\textbf{46.43\%} & \cellcolor[HTML]{E8E5E5}0.00\% & \cellcolor[HTML]{68CBD0}\textbf{43.42\%} & \cellcolor[HTML]{DAE8FC}-6.48\% & 42.43\% & \cellcolor[HTML]{DAE8FC}-8.62\% & \cellcolor[HTML]{68CBD0}11.70\% & \cellcolor[HTML]{E8E5E5}0.00\% & \cellcolor[HTML]{68CBD0}8.49\% & \cellcolor[HTML]{C8DEFE}-27.43\% & \cellcolor[HTML]{68CBD0}5.80\% & \cellcolor[HTML]{BCD6FC}-50.43\% \\
LibPecker & \multicolumn{1}{l}{53.09\%} & \cellcolor[HTML]{E8E5E5}0.00\% & 63.57\% & \cellcolor[HTML]{FFCE93}+19.74\% & 34.81\% & \cellcolor[HTML]{BED6F8}-34.45\% & \cellcolor[HTML]{FE996B}98.72\% & \cellcolor[HTML]{E8E5E5}0.00\% & \cellcolor[HTML]{FE996B}\textbf{95.13\%} & \cellcolor[HTML]{ECF4FF}-3.64\% & \cellcolor[HTML]{FE996B}34.95\% & \cellcolor[HTML]{A0C7FF}-64.60\% \\
ORLIS & \multicolumn{1}{l}{61.16\%} & \cellcolor[HTML]{E8E5E5}0.00\% & 61.69\% & \cellcolor[HTML]{FCE3D4}+0.87\% & 40.39\% & \cellcolor[HTML]{BED6F8}-33.96\% & 63.51\% & \cellcolor[HTML]{E8E5E5}0.00\% & 60.31\% & \cellcolor[HTML]{DAE8FC}-5.04\% & 30.41\% & \cellcolor[HTML]{BCD6FC}-52.12\% \\
LibRadar & \multicolumn{1}{l}{\cellcolor[HTML]{FE996B}\textbf{93.80\%}} & \cellcolor[HTML]{E8E5E5}0.00\% & 94.43\% & \cellcolor[HTML]{FCE3D4}+0.67\% & \cellcolor[HTML]{68CBD0}24.51\% & \cellcolor[HTML]{79ACF5}{\color[HTML]{FFFFFF} \textbf{-73.87\%}} & 65.43\% & \cellcolor[HTML]{E8E5E5}0.00\% & 63.38\% & \cellcolor[HTML]{ECF4FF}-3.13\% & \textbf{8.96\%} & \cellcolor[HTML]{3F7ACF}{\color[HTML]{FFFFFF} \textbf{-86.31\%}} \\
LibScout & \multicolumn{1}{l}{79.11\%} & \cellcolor[HTML]{E8E5E5}0.00\% & \cellcolor[HTML]{FE996B}\textbf{94.93\%} & \cellcolor[HTML]{FFC987}+20.00\% & \cellcolor[HTML]{FE996B}\textbf{98.07\%} & \cellcolor[HTML]{F8A102}\textbf{+23.96\%} & 88.73\% & \cellcolor[HTML]{E8E5E5}0.00\% & 27.40\% & \cellcolor[HTML]{A0C7FF}\textbf{-69.12\%} & 22.02\% & \cellcolor[HTML]{74A6EE}{\color[HTML]{FFFFFF} \textbf{-75.18\%}} \\
\bottomrule[0.9pt]
\end{tabular}
\label{tbl:obf_cmp_tools}
\vspace{-2ex}
\end{table*}

Table~\ref{tbl:obf_cmp_tools} shows the detection results regarding precision and recall. We use the changed rate to indicate the effects of obfuscators on TPL detection rate. The changed rate shows the detection rate difference between apps without obfuscation and the apps with obfuscation. The formula of changed rate is as follows:

\begin{equation}
changed\ rate\ = \frac{X_{obf}-X_{no}}{X_{no}}
\label{equ:changed_rate}
\end{equation}

Where X indicates the detection rate. We first use the detection rate of apps with obfuscation minus the detection rate of the apps without obfuscation to get the difference, and then we use the difference to divide the detection rate of the apps without obfuscation. In this way, we can get the changed rate (the same below).

Considering the precision in TABLE~\ref{tbl:obf_cmp_tools}, we also can find Proguard cannot affect the precision of all tools. Compared the precision of apps without obfuscation, we find the precision of most tools regarding apps obfuscated by Allatori increased except LibID. The results of DashO are totally opposite; the precision of all tools is decreased except LibScout. The reason for the increase in precision is because the decrease in FPs is greater than the decrease in TPs.
For example, if a tool can identify four TPLs in an app without obfuscation, there are three true positives among the four apps. The original precision is 75\%. However, if the app is obfuscated, this tool can only identify one TPL and this TPL is also the TP, the precision becomes to 100\%.  Compared with Allatori, DashO has better obfuscated capability so that it can dramatically decline the TPs. Thus, the precision of most tools is also declined.

As for the recall, it can directly reflect the changes of identified TPLs and the resilient capability of each tool towards different obfuscators. Thus, we mainly refer to recall to analyze the obfuscation-resilient capability of each tool.
According to the TABLE~\ref{tbl:f1-value_tools_obf} \& TABLE~\ref{tbl:obf_cmp_tools}, we can find the same situation, the recall of all tools keep unchanged for apps obfuscated by Proguard. However, they all fail to effectively detect TPLs obfuscated by DashO, leading to a sharp decline in detection rate. 
The recall of all tools has dropped by more than 50\% when handled the apps obfuscated by DashO. Compared with the apps without obfuscation, the recall of LibRadar drops the most, dropping by 86.13\%, followed by LibScout (dropping by 75.18\%)
As we can see, the code obfuscation effects of Allatori lies between Proguard and DashO. Allatori has the most apparent impact on LibScout, the recall decreased 69.12\% and it has the smallest impact on LibRadar, the recall only decreased 3.13\%. 


Based on the result, we can find that all TPL detection tools are obfuscation-resilient to identifier/package renaming.
DashO has the best obfuscation performance and Proguard has the worst obfuscation performance.
LibPecker has the best comprehensive performance while ORLIS is the most effective tool to defend against the three popular obfuscators.

\noindent $\bullet$\textbf{Reason Analysis.} There are two main reasons for such differences in detection rate decline: (1) enabled obfuscation strategies in different obfuscation tools; the more obfuscation strategies are enabled, the lower the detection rate is. (2) Even if apps use the same code obfuscation strategy, different tools have different implementations and final effects may also be different.
In our dataset, we find that the package flattening strategy implemented by Proguard only changes the package name, while both Allatori and DashO change the package name/hierarchy structure and DashO even also includes the class encryption in this process, which directly affects the detection rate. 
It is worth noting that the recall (without obfuscation) of some tools is higher than that in C2 because the apps in C2 are closed-source apps from Google Play, and some of them have been obfuscated by developers.

\subsubsection{\textbf{Evaluation towards Different Obfuscation Techniques}.}
We evaluate the capabilities of the {five} selected TPL detection tools towards defending against three obfuscation techniques: 1) control flow obfuscation, 2) package flattening, 3) dead code removal.
%
Because all tools are resilient to renaming based on the previous section, we do not compare this obfuscation technique in this section.
We aim to investigate how these different code obfuscation techniques affect the detection performance of each tool and which technique has the most prominent effects on these detection tools.

\begin{table}[t]
 	\centering
	\small
	\caption{Comprehensive performance (F1-value) comparison of existing tools regarding the resiliency to different code obfuscation techniques}
	\vspace{-2mm}
\scalebox{0.94}{
	\begin{tabular}{ccccc}
		\toprule[1.2pt]
		\multirow{2}{*}{\textbf{Tool}} & \multirow{2}{*}{\begin{tabular}[c]{@{}c@{}}\textbf{Without}\\ \textbf{obfuscation}\end{tabular}} & \multicolumn{3}{c}{\textbf{With Obfuscation}} \\ \cline{3-5}
		&     & \textbf{CFO}     & \textbf{PKG FLT}  &  \textbf{Code RMV}   \\ \midrule[0.6pt]
LibID & \cellcolor[HTML]{ECF4FF}19.81\% & \cellcolor[HTML]{ECF4FF}0.00\% & \cellcolor[HTML]{ECF4FF}0.17\% & \cellcolor[HTML]{ECF4FF}2.42\% \\
LibPecker & 69.36\% & \cellcolor[HTML]{EFEFEF}\textbf{69.23\%} & 64.47\% & 64.52\% \\
ORLIS & \cellcolor[HTML]{EAE3E3}\textbf{61.09\%} & 56.77\% & \cellcolor[HTML]{EAE3E3}\textbf{61.09\%} & \cellcolor[HTML]{EFEFEF}\textbf{59.81\%} \\
LibRadar & 75.96\% & 64.51\% & 63.84\% & 63.88\% \\
LibScout & 83.37\% & \cellcolor[HTML]{DAE8FC}\textbf{29.81\%} & \cellcolor[HTML]{DAE8FC}\textbf{28.20\%} & \cellcolor[HTML]{DAE8FC}\textbf{28.17\%} \\ 
\bottomrule[1pt]
	\end{tabular}
}	
	\begin{center}\small
	 \color{black}CFO: Control Flow Obfuscation; PKG FLT: Package Flattening; \\Code RMV: Dead Code Removal 
\end{center}
	\label{tbl:F1_obf_diff_tech}
	\vspace{-3ex}
\end{table}

\begin{table*}[]
\centering
\caption{Comparison results of state-of-the-art TPL detection tools regarding the code obfuscation-resilient capability towards different obfuscation techniques}
\begin{tabular}{ccclclccccccc}
\Xhline{1pt}
\multicolumn{1}{c|}{} & \multicolumn{6}{c|}{\textbf{Precision}} & \multicolumn{6}{c}{\textbf{Recall}} \\ \cline{2-13} 
\multicolumn{1}{c|}{} & \multicolumn{2}{c|}{\textbf{CFO}} & \multicolumn{2}{c|}{\textbf{PKG FLT}} & \multicolumn{2}{c|}{\textbf{Code RMV}} & \multicolumn{2}{c|}{\textbf{CFO}} & \multicolumn{2}{c|}{\textbf{PKG FLT}} & \multicolumn{2}{c}{\textbf{Code RMV}} \\ \cline{2-13} 
\multicolumn{1}{c|}{} & \multicolumn{1}{c|}{} & \multicolumn{1}{c|}{} & \multicolumn{1}{c|}{} & \multicolumn{1}{c|}{} & \multicolumn{1}{c|}{} & \multicolumn{1}{c|}{\textbf{changed}} & \multicolumn{1}{c|}{} & \multicolumn{1}{c|}{} & \multicolumn{1}{c|}{} & \multicolumn{1}{c|}{\textbf{changed}} & \multicolumn{1}{c|}{} & \textbf{changed} \\
\multicolumn{1}{c|}{\multirow{-4}{*}{\textbf{Tool}}} & \multicolumn{1}{c|}{\multirow{-2}{*}{\textbf{ratio}}} & \multicolumn{1}{c|}{\multirow{-2}{*}{\textbf{\begin{tabular}[c]{@{}c@{}}changed\\ rate\end{tabular}}}} & \multicolumn{1}{c|}{\multirow{-2}{*}{\textbf{ratio}}} & \multicolumn{1}{c|}{\multirow{-2}{*}{\textbf{\begin{tabular}[c]{@{}c@{}}changed\\ rate\end{tabular}}}} & \multicolumn{1}{c|}{\multirow{-2}{*}{\textbf{ratio}}} & \multicolumn{1}{c|}{\textbf{rate}} & \multicolumn{1}{c|}{\multirow{-2}{*}{\textbf{ratio}}} & \multicolumn{1}{c|}{\multirow{-2}{*}{\textbf{\begin{tabular}[c]{@{}c@{}}changed\\ rate\end{tabular}}}} & \multicolumn{1}{c|}{\multirow{-2}{*}{\textbf{ratio}}} & \multicolumn{1}{c|}{\textbf{rate}} & \multicolumn{1}{c|}{\multirow{-2}{*}{\textbf{ratio}}} & \textbf{rate} \\ \hline
\midrule
LibID & 0.00\% & - & \multicolumn{1}{c}{1.67\%} & - & \multicolumn{1}{c}{7.31\%} & - & 0.00\% & - & 0.09\% & - & 1.45\% & - \\
LibPecker & \multicolumn{1}{l}{60.12\%} & \cellcolor[HTML]{FFCE93}+12.56\% & 57.40\% & \cellcolor[HTML]{FFCE93}+7.47\% & 57.35\% & \cellcolor[HTML]{FFCE93}+7.38\% & \cellcolor[HTML]{FE996B}\textbf{81.61\%} & \cellcolor[HTML]{BFD8FC}-17.49\% & \cellcolor[HTML]{FE996B}\textbf{73.52\%} & \cellcolor[HTML]{BFD8FC}-25.57\% & \cellcolor[HTML]{FE996B}\textbf{73.74\%} & \cellcolor[HTML]{BFD8FC}-25.45\% \\
ORLIS & \multicolumn{1}{l}{54.84\%} & \cellcolor[HTML]{DAE8FC}-6.88\% & 58.89\% & \cellcolor[HTML]{F4F9FF}0.00\% & 59.04\% & \cellcolor[HTML]{ECF4FF}+0.25\% & 58.86\% & \cellcolor[HTML]{E4EFFE}-7.24\% & 63.46\% & \cellcolor[HTML]{ECF4FF}0.00\% & 60.61\% & \cellcolor[HTML]{E4EFFE}-4.49\% \\
LibRadar & \multicolumn{1}{l}{92.03\%} & \cellcolor[HTML]{FBEBD8}+1.85\% & 93.90\% & \cellcolor[HTML]{FDE9D1}+1.33\% & 91.67\% & \cellcolor[HTML]{FBEBD8}-0.20\% & 49.67\% & \cellcolor[HTML]{BFD8FC}-23.31\% & 48.36\% & \cellcolor[HTML]{BFD8FC}-25.34\% & 49.02\% & \cellcolor[HTML]{BFD8FC}-24.42\% \\
LibScout & \multicolumn{1}{l}{\cellcolor[HTML]{FE996B}\textbf{93.90\%}} & \cellcolor[HTML]{FFCE93}+18.23\% & \cellcolor[HTML]{FE996B}\textbf{92.68\%} & \cellcolor[HTML]{FFCE93}+16.70\% & \cellcolor[HTML]{FE996B}\textbf{92.07\%} & \cellcolor[HTML]{FFCE93}\textbf{+15.93\%} & 17.72\% & \cellcolor[HTML]{80B1F9}{\color[HTML]{FFFFFF} \textbf{-79.81\%}} & 16.63\% & \cellcolor[HTML]{80B1F9}{\color[HTML]{FFFFFF} \textbf{-81.05\%}} & 16.63\% & \cellcolor[HTML]{80B1F9}{\color[HTML]{FFFFFF} \textbf{-81.05\%}} \\
\bottomrule[0.9pt]
\end{tabular}
\label{tbl:obf-tech_cmp_pr}
\end{table*}

\noindent $\bullet$\textbf{Result.}  We first compared the comprehensive performance by using F1-value to evaluate these state-of-the-art tools. The specific results are presented in TABLE~\ref{tbl:F1_obf_diff_tech}. 
For apps without obfuscation, LibScout still achieves the best performance, the F1-value is 83.37\%, followed by LibRadar (75.96\%). LibID achieves 19.81\%, which is still the worst. The reasons are the same as before: the performance of LibID is dramatically limited by dex2jar; the code features are too coarse; LibID sets strong constraints for the package structure of TPLs, leading to many false negatives. Because LibID is very picky to the dataset, the result may not precisely and completely reflect the effects of code obfuscation on LibID.

{
The impacts of three obfuscation techniques on the precision and recall of each tool can be seen from TABLE~\ref{tbl:obf-tech_cmp_pr}. 
As for the precision, we can find that most tools have increased. The reasons are the same as that in Section~\ref{eval_obfuscators}. Code obfuscation reduces the number of TPLs that tools can identify. The fuzzy matched TPLs decreased as well.
Besides, we can see the three obfuscation techniques reduce the recall of these tools but different code obfuscation techniques have various effects on different TPL detection tools.
}

{Based on the change rate of recall, the three obfuscation techniques all have great impacts on LibScout, and the recall dropped by about 70\% while they all have minimal impacts on ORLIS, followed by LibPecker.
Three obfuscation techniques have similar effects on LibScout and LibRadar because the changed rate towards the three techniques is close for both LibScout and LibRadar. ORLIS is completely resilient to package flattening, and CFO has greater impacts on ORLIS. 
}


%

For the resiliency to CFO, we can find LibPecker, ORLIS are better than LibRadar and LibScout. Compared with the recall of the apps without obfuscation, ORLIS and LibPecker decreased by 7.24\% and 17.49\%, respectively. LibRadar and LibScout dropped by 23.31\% and 79.81\%.
%

For the resiliency to PKG FLT, ORLIS has a better performance than others, and it is even completely resilient to package flattening. Apart from ORLIS, the recall of remaining tools are dramatically declined, especially LibScout.

The dead code removal can affect all tools. It also dramatically impact on LibScout, and the recall has dropped by about 70\% (decreased more than 80\% of the recall without obfuscation). ORLIS also achieves the best resilience to dead code removal; the decreasing rate is only 4.49\%. The Code RMV has a similar impact on LibPecker and LibRadar; the decreasing rates of recall are 25.45\% and 24.42\%, respectively.


\noindent $\bullet$ \textbf{Reason Analysis.}
{For the control flow randomization, we find rich semantic features that can achieve better resiliency.
LibRadar chooses Android APIs as the code information and LibScout chooses fuzzy method signatures as the code features. These features just include simple syntactic information without any semantic information. In contrast,
LibPecker, LibID, and ORLIS all include the class dependency in their code features and the dependency relationship cannot be changed easily by CFO. Whereas, the code features of LibID are too coarse when it compared with another two tools, leading to many false negatives and false positives. Besides, LibID considers the internal package structures cannot be changed by obfuscation, this assumption is wrong and the package match constraints are too strict, which directly affects its performance.}

{For package flattening, apart from ORLIS, all tools more or less depend on the package structure to identify TPLs. Thus, ORLIS is completely resilient to the package flattening technique. ORLIS has shown that we can only use the class dependency to split TPL candidates from non-primary modules and also can be resilient to package flattening. Package flattening cannot change the class dependency.} 

{Dead code removal can delete the code of a TPL that is not invoked by host apps, which leads to the code of in-app TPLs inconsistent with original TPLs. Without a doubt, that brings a great challenge to TPL detection.}

{Besides, we find the feature granularity also can affect obfuscation-resilient capabilities. LibPecker and ORLIS achieve better performance than other tools in this experiment. LibPecker and ORLIS use fine-grained code features (i.e., class level) that are not sensitive to small code changes. In contrast, LibScout and LibRadar use coarse-grained features (i.e., package level), whose hash values may easily change due to slight code modification.}

\vspace{2mm}

\noindent\fbox{
	\parbox{0.95\linewidth}{
		\textbf{Conclusion to C5:} 
{ORLIS outperforms other tools in defending against different obfuscators and different obfuscation techniques. Considering the overall performance, LibPecker can achieve better performance than others.
LibID is picky to the dataset, and dex2jar dramatically limits its performance. Tools using fine-grained features have better performance than those using coarse-grained features regarding defending against code obfuscation. The code features include more semantic information, which can achieve better resiliency to code obfuscation.
}
}}

\vspace{-2ex}
\subsection{C6: Ease of Use}
\label{sec:easy_of_use}
Whether a tool is user-friendly is an essential factor in evaluating the usability of the tool.
We attempt to compare the usability of the available tools (i.e., LibID, LibPecker, ORLIS, LibRadar, LibD\footnote{We consider the usability of LibD though it was reported containing a calculation error, which would not affect the installation and using process.}, and LibScout) from three aspects: 1) the installation and setup process, 2) the usage steps, and 3) the result presentation.
To assess them objectively, we design a questionnaire \cite{questionnaire}
and recruit participants to rate these tools from the three aspects.

\noindent \textbf{Participant Recruitment.} We recruit {20} people from different industrial companies and universities via word-of-mouth, who are developers in IT companies, post-doc, Ph.D. students, etc.  
To minimize the interference factors due to lack of professional factors of participants, all the participants we recruited have over three years of experience in Android app development, and they are from different countries such as Singapore, Germany, China and India. The participants have diversified backgrounds. 
Besides, they did not install or use these tools before. 
The participants received a \$50 coupon as compensation for their time.

\noindent \textbf{Experiment Procedure.} We provided the links of the source code together with the instruction files (prepared by the tools' authors) that guide participants to install and use these tools. Note that since some tools require users to take apks (and TPLs) as input, we also provide some sample apps and sample TPLs in case that participants have no idea about where to download the input data, which may hinder the process of using them. {The samples we select from the second dataset for C2, to ensure all tools can run these samples successfully. We have published this dataset, and you can download it from our website~\cite{our_lib}.}
We ask the participants to install and use these six tools one by one, and rate each tool from the aforementioned three aspects. The specific rating criteria can be seen in Table~\ref{tbl:rating_criteria}. We offer the same device with the same running experiment for all participants. 
They all carried out experiments independently without any discussions with each other and they were encouraged to write some comments about each tool. After finishing the tasks, we also interviewed them about the user experience with detailed records. The demographics information of the participants and detailed survey and investigation results can be found from our website~\cite{our_lib}.

\begin{table}[t]
\centering
\small
\caption{Rating options for each item in the questionnaire}
\vspace{-1mm}
\scalebox{0.72}{
\begin{tabular}{r|
>{\columncolor[HTML]{FB9C52}}c |
>{\columncolor[HTML]{FDB781}}c |
>{\columncolor[HTML]{FFE9D8}}c |c}
\cline{2-5}
\textbf{Installation} & {\color[HTML]{000000} \begin{tabular}[c]{@{}c@{}}Easy\\ $\bigstar\bigstar\bigstar\bigstar$\end{tabular}} & \begin{tabular}[c]{@{}c@{}}Acceptable\\ $\bigstar\bigstar\bigstar$\end{tabular} & \begin{tabular}[c]{@{}c@{}}Complicated\\ $\bigstar\bigstar$\end{tabular} & \multicolumn{1}{c|}{\cellcolor[HTML]{FDF6F1}\begin{tabular}[c]{@{}c@{}}Very  complicated\\ $\bigstar$\end{tabular}} \\ \hline
\textbf{Usage} & {\color[HTML]{000000} \begin{tabular}[c]{@{}c@{}}Easy\\ $\bigstar\bigstar\bigstar\bigstar$\end{tabular}} & {\color[HTML]{000000} \begin{tabular}[c]{@{}c@{}}Acceptable\\ $\bigstar\bigstar\bigstar$\end{tabular}} & {\color[HTML]{000000} \begin{tabular}[c]{@{}c@{}}Complicated\\ $\bigstar\bigstar$\end{tabular}} & \multicolumn{1}{c|}{\cellcolor[HTML]{FDF6F1}{\color[HTML]{000000} \begin{tabular}[c]{@{}c@{}}Very  complicated\\ $\bigstar$\end{tabular}}} \\ \hline
\textbf{Output} & {\color[HTML]{000000} \begin{tabular}[c]{@{}c@{}}Clear \& direct\\ $\bigstar\bigstar\bigstar$\end{tabular}} & {\color[HTML]{000000} \begin{tabular}[c]{@{}c@{}}Understand/ \\ not concise\\ $\bigstar\bigstar$\end{tabular}} & {\color[HTML]{000000} \begin{tabular}[c]{@{}c@{}}Confusing \\ $\bigstar$\end{tabular}} & \multicolumn{1}{l}{} \\ \cline{2-4}
\end{tabular}%
}
\label{tbl:rating_criteria}
\vspace{-2ex}
\end{table}

\begin{figure}[t]
	\centering
	\includegraphics[scale=0.48]{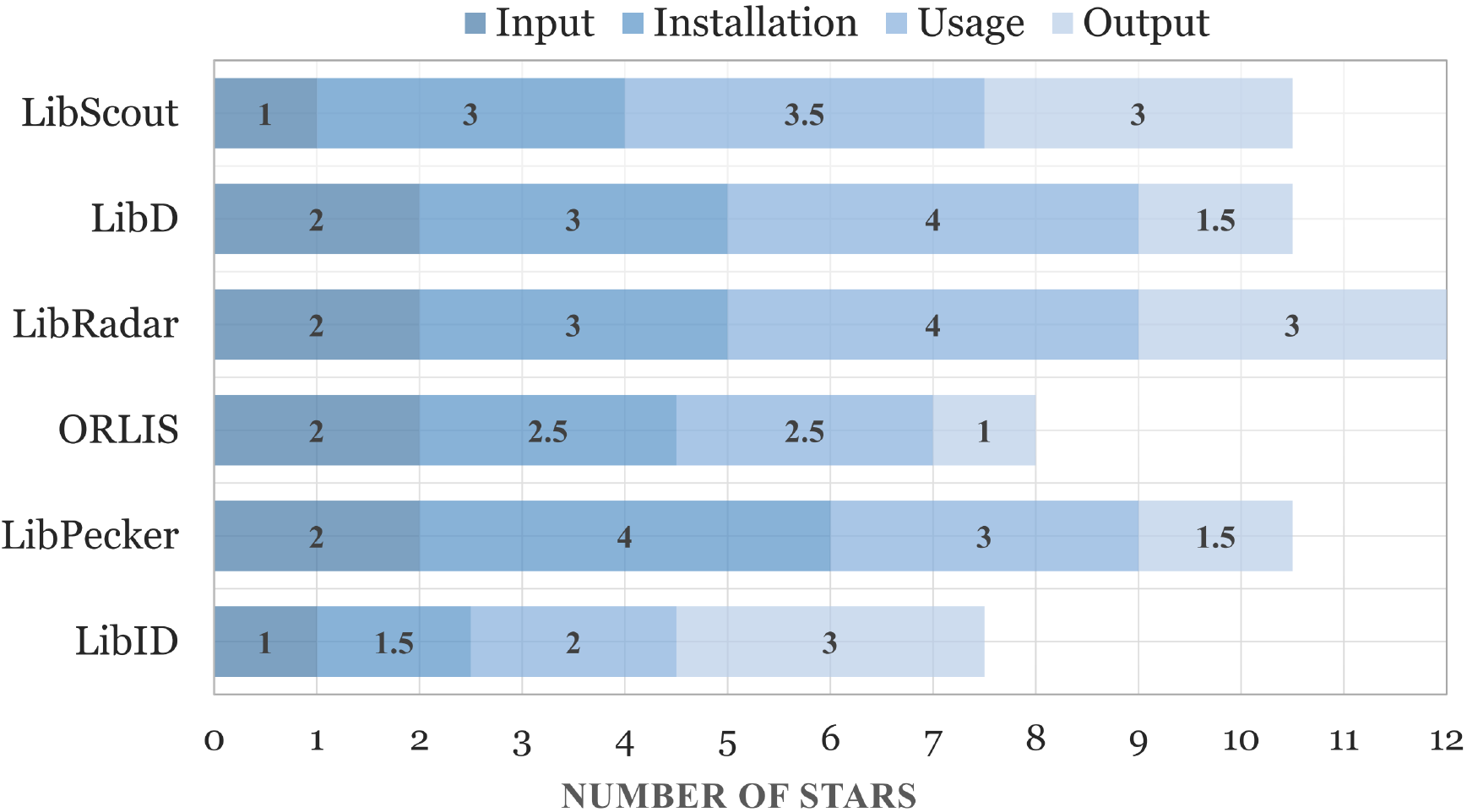}\\
	\vspace{-2mm}
	\caption{ Rating result of each tool from the questionnaire}
	\label{fig:survey}	
	\vspace{-2ex}
\end{figure}

\noindent \textbf{Results.} 
Fig.~\ref{fig:survey} shows the results of the questionnaire. For each rating item, we take the average of the rating stars from all participants. According to Figure~\ref{fig:survey}, we can find that LibRadar gets the most stars while ORLIS receives the lowest score. 

\smallskip
\noindent $\bullet$ \textbf{Installation:}
As for the installation process, LibPecker gets the highest score (4 stars) since it only needs one command, and LibID is regarded to be the most complicated one because it requires participants to install the Gurobi Optimizer~\cite{gurobi} and register the license by themselves.
Most users commented that it is difficult to follow the instructions of Gurobi Optimizer, which wastes them much time.
%
%
%
%
\noindent Both LibRadar and LibD get 3 scores because they just need to install basic Java and python running environments and then users can run them.
The installation of ORLIS and LibScout is almost acceptable (2.5 stars). Both of them require users to download the Android SDK. Besides, ORLIS requires users to download some dependencies and TPLs to make it run.

\smallskip
\noindent $\bullet$ \textbf{Usage:}
As for the usage, LibRadar and LibD get the highest scores, while LibID is regarded as the most unsatisfactory tool. The results show that participants consider more about the execution efficiency of the tools when using them. They are reluctant to wait too much during the investigation.

\smallskip
\noindent $\bullet$ \textbf{Output:} 
According to the results, participants thought the detection results of LibID, LibRadar, and LibScout are much easier to understand; and ORLIS gets the lowest score since it only provides the matching relations of the class name, without telling the apps or TPLs that the class belongs to, making participants confused.

\vspace{0.05mm}
\noindent ``\textit The results of LibID, LibRadar and LibScout are easy to understand. All of the results are represented in ``.json'' format, I can quickly find the in-app TPLs, the similarity value and other meta information.''
\vspace{0.05mm}

\noindent The result provided by LibD is the MD5 of TPLs, which requires users to find the used TPLs by mapping with the database file provided by LibD:

\vspace{0.05mm}
\noindent ``\textit{I can understand the result of LibD, but not very direct and clear. Some information that I really interested in is missing, e.g., {the similarity value} and library name.}''

\vspace{1mm}
\noindent\fbox{
	\parbox{0.95\linewidth}{
		\textbf{Conclusion to C6:} 
        LibRadar gets the highest score from participants mainly due to its simple usage and user-friendly output format, which is regarded as the most easy-to-use tool. ORLIS should be improved most, especially the result representation.
	}}

\section{Threats To Validity}
\label{sec:threats}

The validity of this work may be subject to the following threats.

\noindent$\bullet$ \textbf{Paper Collection}.  Even if we have followed a well-defined systematic literature review methodology
to collect papers on Android TPL detection, it is still possible to leave out some publications. To reduce the threat, we first conduct a backward-snowballing on the remaining papers to mitigate the omission. We then reviewed all the references of our collected papers and tried to find the omission papers. We define the keywords to search the relevant papers, but these keywords cannot cover all papers. Our paper set does not include the books, Master's, or Ph.D. dissertations. To mitigate this threat, we searched these authors' articles related to TPL identification published in conferences and journals.

\noindent$\bullet$ \textbf{Errors in manual verification.} This work involved much manual verification (e.g., identifying the TPL-app mapping information). Unfortunately, such a manual process cannot guarantee the results with 100\% accuracy. To mitigate this threat, we have cross-validated all results for consistency.

\vspace{-1ex}
\section{Tool Enhancement}
 Based on our study, we fixed the defects of existing TPL detection tools with: (1) ability to handle all formats of TPLs. {We modify the code of LibID, LibPecker, and ORLIS to let them can handle TPL in ``.aar'' format. We first identify the format of each TPL, and when we find the ``.aar'' file, we unzip this file and get the ``.jar'' file.}
 With the same precision, the recall of LibID, LibPecker, and ORLIS have increased by 6.30\%, 19.32\%, and 12.59\%, respectively.
(2) We modify the code of LibRadar and ORLIS to let them can handle APKs ($API>21$) that include the multi-dex. The original code just read only single ``.dex'' file; we read the list of the ``.dex'' files as the input and send the result to the next process. {With the same precision,}
the recall of LibRadar and ORLIS has increased by 16.15\% and 14.69\%. (3)
Apart from the aforementioned common problems, we find that LibID can be enhanced from two aspects: 1) Bypass reverse-engineering protection (for dex2jar tool), and Android developers can prevent its Java code from being converted by dex2jar by inserting the syntax error of bytecode into the java function that they want to protect. By exploiting the syntax check mechanism of dex2jar, developers can achieve the purpose of dex2jar conversion error. We fixed this problem by modifying the \texttt{TypeClass.java} file to ignore the type errors if the apps are protected in this way. 
2) fix run-time errors. Users directly use the commands, which will meet a crash when the database contains large batches of TPL files. We enhance LibID by splitting the signature volume into smaller units (i.e., 100 TPLs as a unit) and merge the detection results.
The detection rate of LibID increased by 14.84\%.
We have published the related code on GitHub~\cite{LibID_us} anonymously. Users can directly use our enhanced tools to achieve better performance.  
For more details, please refer to our website~\cite{our_lib}. 


\noindent\textbf{Online service for TPL detection.} 
To make it more convenient for users to access and compare these tools, we build a framework which integrates the five publicly available tools and make it as an online service~{\cite{our_lib}} to detect the in-app TPLs. Our framework can easily be extended if new tools are available. Users can upload an app to the online platform, and our framework will show the detection result of each tool. Users can compare each tool intuitively and clearly observe the commonalities and differences of the results.

\vspace{-2ex}
\section{Discussion}
\label{sec:discussion}

Based on our evaluation, we highlight lessons learned from different perspectives,
and give implications for tool enhancement and provide insights for future research.

\vspace{-2ex}
\subsection{Lessons Learned} 

\subsubsection{\textbf{Application Scenarios}}
We give an in-depth discussion on existing Android TPL detection tools and propose tool selection suggestions for different stakeholders with different purposes.
\begin{itemize}

\item  For \textbf{malware/repackaged app detection}, we suggest choosing the LibSift. Since the main purpose of malware/repackaged app detection is to filter TPLs out, we do not need to choose some tools that can identify the specific TPL versions; these tools are not cost-effective. LibSift just can identify the host app and TPLs by using an effective module decoupling algorithm (cf. Sec 4).

\item For \textbf{vulnerable in-app TPL detection}, we recommend LibID that has better performance in identifying the specific library versions in TPL detection. We can use LibID to find the in-app library versions and confirm whether these versions include the vulnerabilities from some public vulnerability database like NVD~\cite{NVD} (cf. Sec 6.4).

\item For \textbf{component analysis of apps}, we recommend ORLIS or LibScout. If an app includes many obfuscated TPLs, we recommend ORLIS that is proven to have the best code obfuscation-resilient capability against common obfuscators and common obfuscation techniques. Otherwise, LibScout is a good choice; it can achieve high precision and recall compared with other tools (cf. Sec 6.2).

\item For \textbf{large-scale TPL detection}, we suggest using LibRadar that has high-efficiency (i.e., 5.53s per app on average). It is suitable for large-scale detection (cf Sec 6.3). 

\item For \textbf{advertising investors or developers who want to choose popular ad networking (i.e., ad libraries)} to show their ads, we also recommend them to use LibRadar that can efficiently find commonly-used ad libraries in Android apps at market scale. 
With these identified TPLs, developers can choose some competitive ones to embed in their products to ensure their competitive edge in the market.

\end{itemize}

\subsubsection{\textbf{Limitations of Existing Tools.}}

In this section, we present the limitations of previous research to raise the attention to future tool implementation. The limitations are mainly reflected in three aspects.

\noindent $\bullet$ \textbf{Ignore New Features.} Android system updates frequently and usually introduces new features. However, current approaches seem to pay less attention to these new features, directly compromising the detection performance. For example, (1) \textit{Android 64K limit problem}~\cite{ART}, Since Android 5.0, it supports compiling an app into multiple Dex files if it exceeds the limited size. However, LibRadar and ORLIS ignored this feature and read only one ``.dex'' file to generate the TPL feature, leading to loss of TPL features and low recall.
(2) \textit{New app formats.} 
The apps published with Android App Bundle \cite{aab} format will finally be released with the file suffix ``.apks'' or ``.xapk'', which cannot be directly handled by all existing tools.
(3) \textit{Ignore attributes of TPLs.}
Some tools even ignored some basic attributes of TPLs, but handling these attributes is not very challenging and also can help improve performance dramatically. 
For example, LibID, LibPecker, and ORLIS only can handle the TPLs in ``jar'' format and ignore the TPLs in ``aar'' format, which can directly lead to false negatives because the TPL signature database lacks these version files. Whereas,
``aar'' format TPL is a kind of exclusive Android TPL; these TPLs are wildly-used by many app developers.

\noindent $\bullet$ \textbf{Set strong assumptions.} Some tools set too strong or even inaccurate assumptions, which directly affects their performance.
Most tools (e.g., \textbf{LibRoad, LibRadar, LibScout}) roughly treat each package tree as TPL instance. Apart from ORLIS, existing tools more or less depend on package structure to split TPL candidates and identify TPLs based on package structure/name.  In fact, the package structure is fragile and unreliable based on our analysis. We should avoid choosing this feature in TPL identification. 
\textbf{LibPecker} assumes that the package structure is preserved during obfuscation. \textbf{LibID} assumes that inter-package structures will not change during obfuscation. However, these assumptions are not accurate, which directly affects their recall.

\noindent $\bullet$ \textbf{Be Constrain by selected tools.} \textbf{LibID} is dramatically limited by dex2jar (cf. Section\ref{eval_obfuscators}).

\vspace{-1ex}
\subsection{Implications} 
Via our study, we present our takeaways of TPL detection. 

\noindent$\bullet$ \textbf{Selection of reverse-engineering tool}, we suggest that researchers choose tools with high stability and good compatibility with Android and Java, such as \textit{Soot} and \textit{Androguard}.

\noindent$\bullet$ \textbf{Module Decoupling.} We find that the module decoupling features including the package structure are error-prone; we do not recommend researchers adopt package hierarchy structure in TPL candidate construction. We suggest considering the strategy of ORLIS that uses class dependency as the module decoupling feature, which has proved its effectiveness and can achieve the best resiliency towards package flattening obfuscation and package mutation. Moreover, this method can avoid separating a complete TPL that depends on other TPLs into several parts. 

\noindent$\bullet$ \textbf{Feature Extraction.} We find that if the features include rich semantic information, the tool can achieve better resiliency to code obfuscation.
Researchers also can consider the class dependency. We find that when the class dependencies are encoded into code feature, it usually can achieve better resiliency to code shrink. Research should consider the trade-off of the feature granularity. The finer-grained features can achieve better performance in version identification but also can cost more time.

\noindent$\bullet$ \textbf{Comparison Strategy.} We suggest developers can leverage the strategy that is adopted by OSSPoLICE and LibRoad. We can consider using different strategies to collapsed packages and non-collapsed packages. For non-obfuscated packages, we can use package names to narrow down the search space. 
It can first compare the coarse-grained features and find the potential TPL candidate and  
then compare the fine-grained features, which can improve the efficiency.

\vspace{-1ex}
\subsection{Future Research Directions}
We highlight some insights to inspire and motivate future research on TPL detection techniques.

(1) \textbf{Consider TPLs developed in other languages.} According to our study, we find that over 15\% in-app TPLs are developed in Kotlin~\cite{Kotlin}.
However, some grammar rules of Kotlin are different from Java, which can directly affect the performance of existing TPL detection tools. 
Specifically, the source files of Kotlin can be placed in any directory, which can cause similar effects like the package flattening obfuscation technique. Therefore, existing tools depending on package structure to generate code features may become ineffective if app developers customize Kotlin TPLs, especially modify the package name/structure that can easily change the signatures of TPLs.
Besides, future researchers also can consider the native library (C/C++) detection and related security problems understanding. 

(2) \textbf{Detect vulnerable TPLs.} Although LibScout claims that it can detect vulnerable TPLs, the complete dataset of vulnerable TPLs is missing now. We know little about the risks of vulnerable TPLs and infected apps. Future research on vulnerable TPL detection and understandings is necessary and meaningful.

(3) \textbf{Catch emerging TPLs.} Existing tools rely on a reference database to identify TPLs, which limits their ability to detect TPLs in the database, thus cannot identify newly-published TPLs that are not in the database. 

(4) \textbf{Identify TPLs by using dynamic techniques.} Current methods for TPL detection are static analysis, which cannot identify TPLs with dynamic behaviors. For example, some TPLs can be dynamically loaded at run-time. Some TPLs can leverage the reflection and dynamic classload techniques to hide some code, which is difficult to find by using these static analysis methods.

\section{Conclusion}
\label{sec:conclusion}

In this paper, we investigated existing TPL detection techniques from both literature-based perspectives and practice-based perspectives. We conducted a thorough comparison on existing tools from six aspects, including TPL instance construction accuracy, effectiveness, efficiency, version identification code obfuscation-resilience capability, and ease of use. We summarized useful insights and revealed the advantages and disadvantages of these state-of-the-art TPL detection tools. 
We discussed lessons learned from different perspectives, and enhanced existing tools and further provided an online service for TPL detection. 
Besides, our dataset and evaluation details are publicly available.
We believe our research can provide the community with a clear viewpoint on this direction, help developers to develop better tools and inspire future researchers to find more creative ideas in this area.




\footnotesize
\bibliographystyle{IEEEtran}
\bibliography{Libdetect}

\end{document}